\documentclass{optica-article}

\journal{opticajournal} 

\articletype{Research Article}
\usepackage{multirow}

\usepackage{lineno}

\newcommand{\tMZI}{\tau_{\textrm{MZI}}}
\newcommand{\tPLL}{\tau_{\textrm{PLL}}}

\newcommand{\sPLL}{\sigma_{\textrm{PLL}}}
\newcommand{\sigit}{\sigma_{\textrm{iteration}}}
\newcommand{\tmix}{\tau_{\textrm{mix}}}

\newcommand{\tEOM}{\tau_\mathrm{EOM}}
\newcommand{\FEOM}{F_\mathrm{EOM}}
\newcommand{\FFF}{F_\mathrm{FF}}
\newcommand{\FMZI}{F_\mathrm{MZI}}
\newcommand{\tFF}{\tau_\mathrm{FF}}
\newcommand{\fAOM}{f_\mathrm{MZI}}
\newcommand{\dtEOM}{\delta\tau_{\mathrm{EOM}}}

\begin{document}

\title{Versatile, fast and accurate frequency excursions with a semiconductor laser}

\author{Thomas Llauze,\authormark{1} Félix Montjovet-Basset,\authormark{2} and Anne Louchet-Chauvet\authormark{1,*}}

\address{\authormark{1}ESPCI Paris, Université PSL, CNRS, Institut Langevin, 75005 Paris, France\\
\authormark{2}Chimie ParisTech, PSL University, CNRS, Institut de Recherche de Chimie Paris, Paris, 75005, France\\}

\email{\authormark{*}anne.louchet-chauvet@espci.fr} 


\begin{abstract*}
Achieving accurate arbitrary frequency excursions with a laser can be quite a technical challenge, especially when steep slopes (GHz/$\mu$s) are required, due to both deterministic and stochastic frequency fluctuations. In this work we present a multi-stage correction combining four techniques: pre-distorsion of the laser modulation, iterative correction, opto-electronic feedback loop and feed-forward correction. This combination allows not only to compensate for the non-instantaneous response of the laser to an input modulation, but also to correct in real time the stochastic frequency fluctuations. We implement this multi-stage architecture on a commercial DBR laser and verify its efficiency, first with monochromatic operation and second with highly demanding frequency excursions. We demonstrate that our multi-stage correction not only enables a strong reduction of the laser linewidth, but also allows steep frequency excursions with a relative RMS frequency error well below $1$\%, and a laser spectral purity consistently better than $100$~kHz even in the midst of GHz-scale frequency excursions.
\end{abstract*}

\section{Introduction}

The ability to rapidly and precisely tune the frequency of a laser, known as "agility", is key for a large number of applications like FMCW lidar~\cite{Gazizov2022lowpixel,zhang2019laser}, rapid wavelength switching in telecommunications~\cite{kranendonk2005modeless}, wideband optical signal processing~\cite{babbitt2014spectral} and quantum applications~\cite{kinos2021roadmap}. 

The simplest approach to generate fast laser frequency chirps is to directly modulate the wavelength of a laser. Extended cavity diode lasers (ECDL) are widely tunable, but this tunability relies on the mechanical movement of a grating or mirror, which limits both reproducibility and tuning speeds to a few hundred Hz or below. The addition of an intra-cavity electro-optical element has been proposed to overcome these limitations~\cite{levin2002mode}, but the modulation bandwidth is then constrained by its piezoelectric resonances~\cite{crozatier2006phase}. Conversely, silicon nitride-based photonic integrated circuits enable impressive chirp rates in the THz/µs range~\cite{li2022integrated} with up to $500$~MHz modulation frequencies, but with major chirp non-linearity. Finally, monolithic semiconductor lasers (such as DBR or DFB architectures) allow for extended continuous tuning range via the external control of their injection current~\cite{nagarajan1999high,westbrook1984monolithic}, while the corresponding modulation bandwidth can be as high as $100$~MHz~\cite{pan1989modulation} thanks to the very short carrier lifetime in such structures~\cite{zheng2007determination}. Nevertheless, the precision of the frequency excursions decreases with their steepness.

Fast frequency chirps can also be created from any monochromatic laser by single sideband external phase modulation~\cite{yi2021frequency}, which offers ideal modulation linearity and control of the frequency and amplitude of the optical signal waveform. Serrodyne driving of an external phase modulator in the path of a monochromatic laser beam is also an efficient way to achieve precise frequency shifting~\cite{laroche2008serrodyne}, but its implementation at the GHz level imparts severe requirements on the arbitrary waveform generation~\cite{poberezhskiy2005serrodyne}. In any case, for all external modulation  techniques, jitter and frequency noise of the seed laser are likely to lead to stochastic frequency fluctuations that eventually impair the sweep linearity.

Therefore, it appears necessary to deploy versatile and effective tools to enhance the linearity and precision of fast laser frequency scans. Several complementary methods have already been proposed. The pre-distorsion of the command sent to the laser source, based on prior measurement of the modulation input transfer function, corrects the deviations caused by the non-instantaneous response of the laser~\cite{satyan2009precise,kervella2014laser,lihachev2023frequency}. An iterative correction allows to compensate for the frequency deviations due to the nonlinear part of the laser's response~\cite{kervella2014laser,li2020nonlinear,cao2021highly,lihachev2023frequency}. Finally, an improvement in the chirps precision is possible through real-time feedback  correction~\cite{crozatier2006phase, satyan2009precise,kervella2014laser,qin2015coherence} that addresses the stochastic frequency noise associated either with the intrinsic laser or triggered by the frequency sweep. Interestingly, these three methods are not specific to a particular laser design.
However, despite impressively high chirp linearity demonstrated using such correction methods~\cite{qin2015coherence,cao2021highly,lihachev2023frequency}, their operational range is limited to modulation frequencies below a few hundred kHz, typically. This limitation renders them compatible only with relatively slow chirps or rather gentle slope changes.

In this paper, we propose a multi-stage method to achieve precise chirps at the GHz/$\mu$s scale. The method encompasses (i) a pre-distorsion followed by (ii) an iterative optimization of the voltage command used to achieve the frequency excursions, a (iii) opto-electronic feedback loop and (iv) a feed-forward correction of the instantaneous laser frequency. While each of these stages has been used previously to linearize frequency chirps or reduce laser linewidth, as evidenced by prior works~\cite{roos2009ultrabroadband, cao2021highly, li2022linear, greiner1998laser, lintz2017note, cheng2022feed, qin2015coherence}, our novel contribution lies in the comprehensive combination of these four techniques allowing a robust and versatile operation, capable of  addressing both monochromatic operation and arbitrary frequency excursions with steep chirps.

In section~\ref{sec:systematic}, we briefly present the twofold correction method designed to compensate for the systematic deviations to the frequency excursion, including the pre-distorsion based on the laser transfer function, and the iterative correction stage. This method relies on a non-ambiguous measurement of the laser instantaneous frequency. Section~\ref{sec:suppression} is devoted to the description of the real-time correction stage addressing the stochastic deviations to the intended frequency excursion. This correction includes a feedback loop and a feed-forward correction. We demonstrate the potential of this multi-stage architecture by implementing triangular chirps with various amplitudes and modulation frequencies, and an arbitrary frequency excursion involving alternating steep frequency variations and monochromatic operation. We show that a relative RMS frequency error better than $1$\% of the total excursion range can be achieved, including the abrupt slope changes. This demonstration utilizes a commercial DBR laser but could be transposed to any rapidly tunable laser.

\section{Correcting systematic laser frequency deviations}
\label{sec:systematic}

\subsection{Generating and measuring fast frequency excursion}
\label{sec:mzi}

Modulating the frequency of a semiconductor laser by varying the bias current in the laser's gain section leads to strong imperfections, primarily stemming from the non-instantaneous nature of the laser reaction to changes in its bias current. These imperfections can manifest as unwanted non-linearities in the frequency chirps~\cite{zhang2019research}. In order to quantify and correct such imperfections, an access to the laser's instantaneous frequency variations is required. The quality of the frequency excursion will then be estimated by measuring the RMS frequency error with respect to the desired frequency excursion. An interesting figure of merit is the relative RMS, defined as the ratio between this RMS frequency error and the total excursion frequency range.

\begin{figure}[t]
\centering \includegraphics[width=8.5cm]{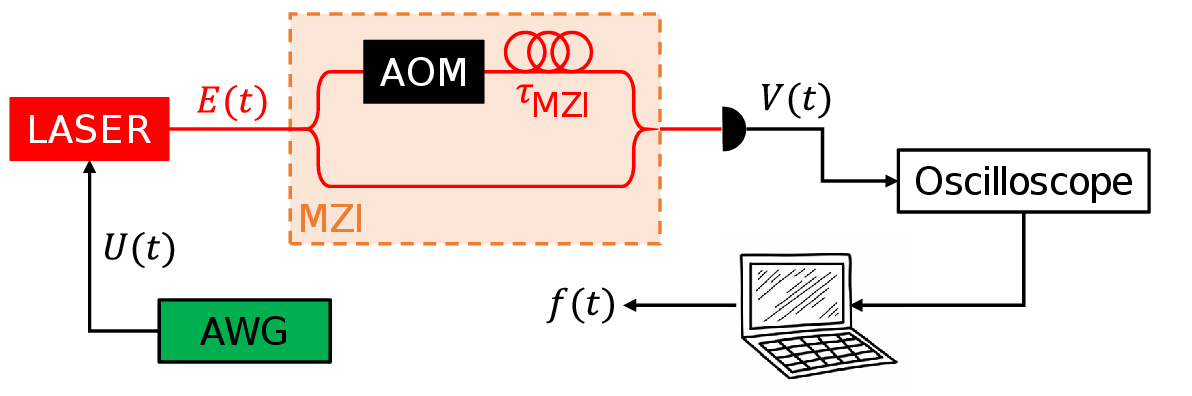}
\caption{Optical setup for instantaneous laser frequency measurement. An arbitrary waveform generator (AWG) issues the voltage command $U(t)$ sent to the laser source. The laser field $E(t)$ enters an unbalanced Mach-Zehnder interferometer (MZI) containing an acousto-optic modulator (AOM) inducing a frequency shift in its long arm. The beatnote $V(t)$ collected on the photodiode is digitized and processed to retrieve the instantaneous laser frequency $f(t)$.}
\label{fig:mzi}
\end{figure}

The instantaneous frequency can be obtained using an unbalanced Mach-Zehnder interferometer (MZI) with an optical delay $\tMZI$ and a frequency shift $\fAOM$ in one of its arms (see Figure~\ref{fig:mzi}). We finally collect the beatnote signal on a photodiode at the output of the MZI. For an incoming laser field $E(t)=E_0 e^{i(2\pi \nu_0 t + \phi(t))}$, $\phi(t)$ being the instantaneous phase of the laser, the beatnote signal at the output of the MZI reads as:
\begin{equation}
    V(t) \propto  1 + \sin{\left( 2 \pi \fAOM t + \phi(t-\tMZI) - \phi(t) \right)}
\end{equation}
The instantaneous frequency $f(t)$, defined as $f(t) = \frac{1}{2\pi} \frac{d\Phi(t)}{dt}$, may then be retrieved from $V(t)$ using the Hilbert transformation~\cite{ahn2007analysis}.

\subsection{Distorted laser response and corresponding correction}
When the laser frequency is modulated via a modulation input on its current supply, a voltage command $U_0(t)$ leads to a distorted frequency excursion $f(t)$. Assuming the laser and current supply operate linearly, this frequency excursion is given by the convolution product of the voltage command and the impulse response $R(t)$ of the system:
\begin{equation}
    f(t) = U_0(t) \otimes R(t)
    \label{eq:transferfunc}
\end{equation}
To make the laser operate according to the user's requirement, one can fully characterize the laser and driver's impulse response $R(t)$ (or, equivalently, its transfer function $H(f)$ expressed in Hz$/$V), and consequently operate the laser with a pre-distorted voltage command $U_{pd}(t)$ given by:
\begin{equation}
	U_{pd}(t) = {\mathscr{F}}^{-1}\left[\frac{\mathscr{F}[f_C(t)](f)}{H(f)}\right]
     \label{eq:precorrection}
\end{equation}
where $f_C(t)$ is the desired frequency excursion  and ${\mathscr{F}}$ is the Fourier transform operator.

In the linear regime, such a pre-distortion ideally compensates for reproducible laser frequency deviations originating from the non-instantaneous response of the laser and driver. However, residual errors may remain due to the effective non-linear response the laser frequency to the injection current~\cite{funabashi2004recent}. To address this, an efficient method consists in iteratively modifying the voltage command with a small correction inferred from the previously measured frequency error~\cite{kervella2014laser,lihachev2023frequency}. More specifically, the $i$-th voltage command is obtained as:
\begin{equation}
	U_i(t) =  {\mathscr{F}}^{-1}\left[\frac{ [\mathscr{F}[U_{i-1}(t)](f) H(f) + \alpha \mathscr{F}[\epsilon_{i-1}(t)](f)}{H(f)}\right],
 \label{eq:iteration}
\end{equation}
$U_{i-1}(t)$ being the previous voltage command, $\epsilon_{i-1}(t)$ the previous frequency error with respect to the desired frequency excursion, and $\alpha$ a numerical parameter chosen between $0$ and $1$ to ensure the convergence.

\subsection{Experimental realisation}
\label{sec:exp1}
We implement the aforementioned double correction scheme on a Photodigm PH795DBR laser operating at $793.4$~nm. Voltage commands are sent to the fast modulation input of the laser current driver (Vescent D2-105), therefore directly addressing the gain section. The successive voltage commands are generated with an arbitrary waveform generator (Tektronix AWG5004). The power variation associated to such a modulation is below 1\%. 
The fully fibered, unbalanced MZI comprises an optical delay $\tMZI = 60$~ns and a $80$~MHz AOM in the same arm. The beatnote signal is collected on a $150$ MHz bandwidth photodiode, sampled with a digital oscilloscope and transferred to a computer to derive the instantaneous frequency of the laser $f(t)$~\cite{ahn2007analysis}.

\begin{figure}[t]
\centering \includegraphics[width=8.5cm]{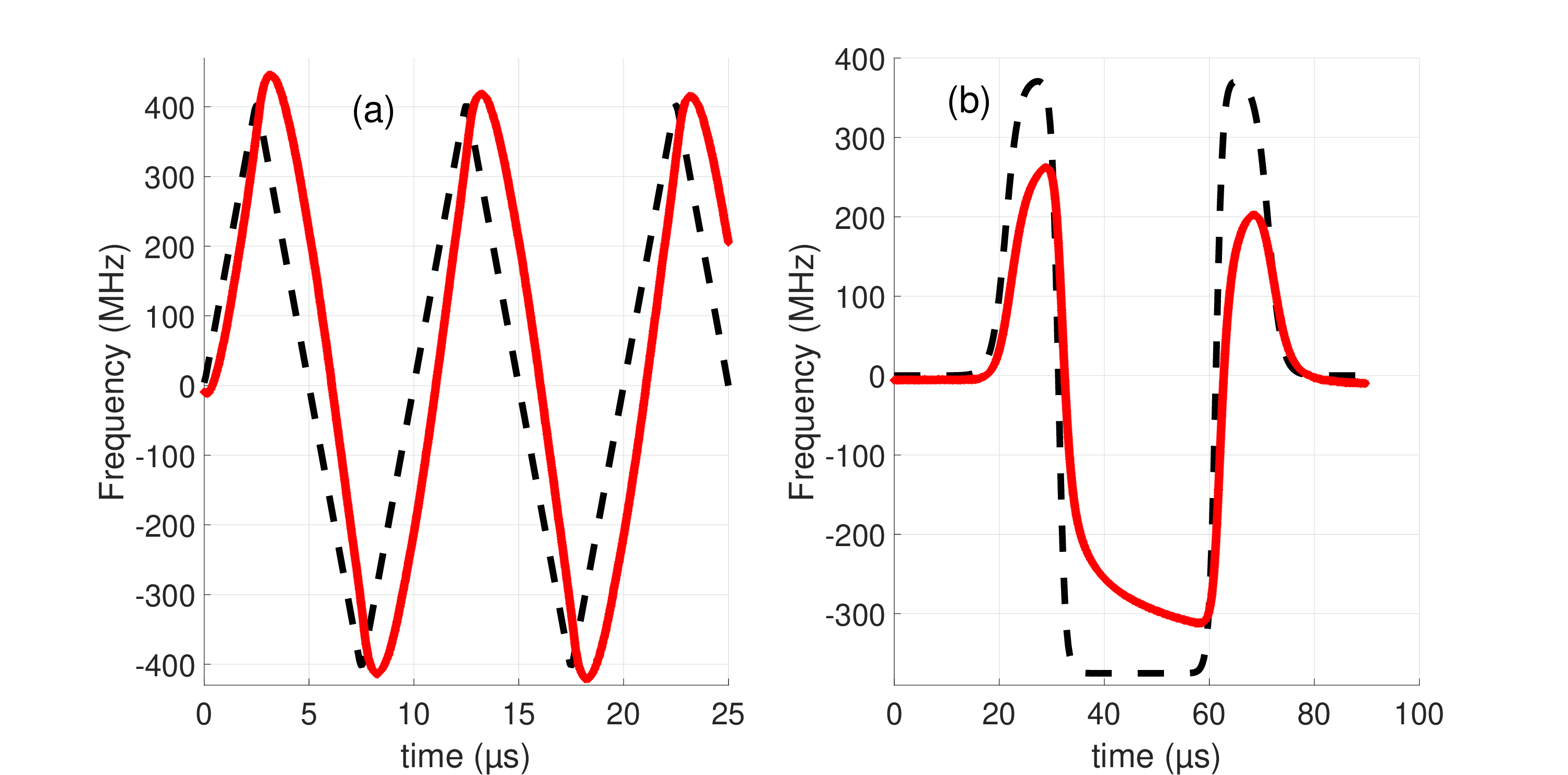}
\caption{Evidence of the non-instantaneous laser response (solid red line) to a rapidly varying drive voltage (black dashed line), in the case of (a) triangular and (b) arbitrary command.}
\label{fig:nocorrection}
\end{figure}

Two types of excursion are tested: first, a periodic, triangular frequency excursion comprised of two opposite chirps, with a $800$~MHz chirp span and $100$~kHz modulation frequency. Second, an arbitrary excursion with two opposite $2.5~\mu$s-long chirps over a $750$~MHz range, corresponding to a $300~$MHz/$\mu$s chirp rate, separated by a $20~\mu$s monochromatic operation interval (see Figure~\ref{fig:nocorrection}). The drive voltage and the corresponding laser response show a strong distortion, with frequency errors as high as $300$~MHz. In the triangular case, we measure a $72.9$~MHz RMS frequency error, \emph{ie} a $9.1\%$ relative RMS.

\begin{figure}[t]
\centering \includegraphics[width=8.5cm]{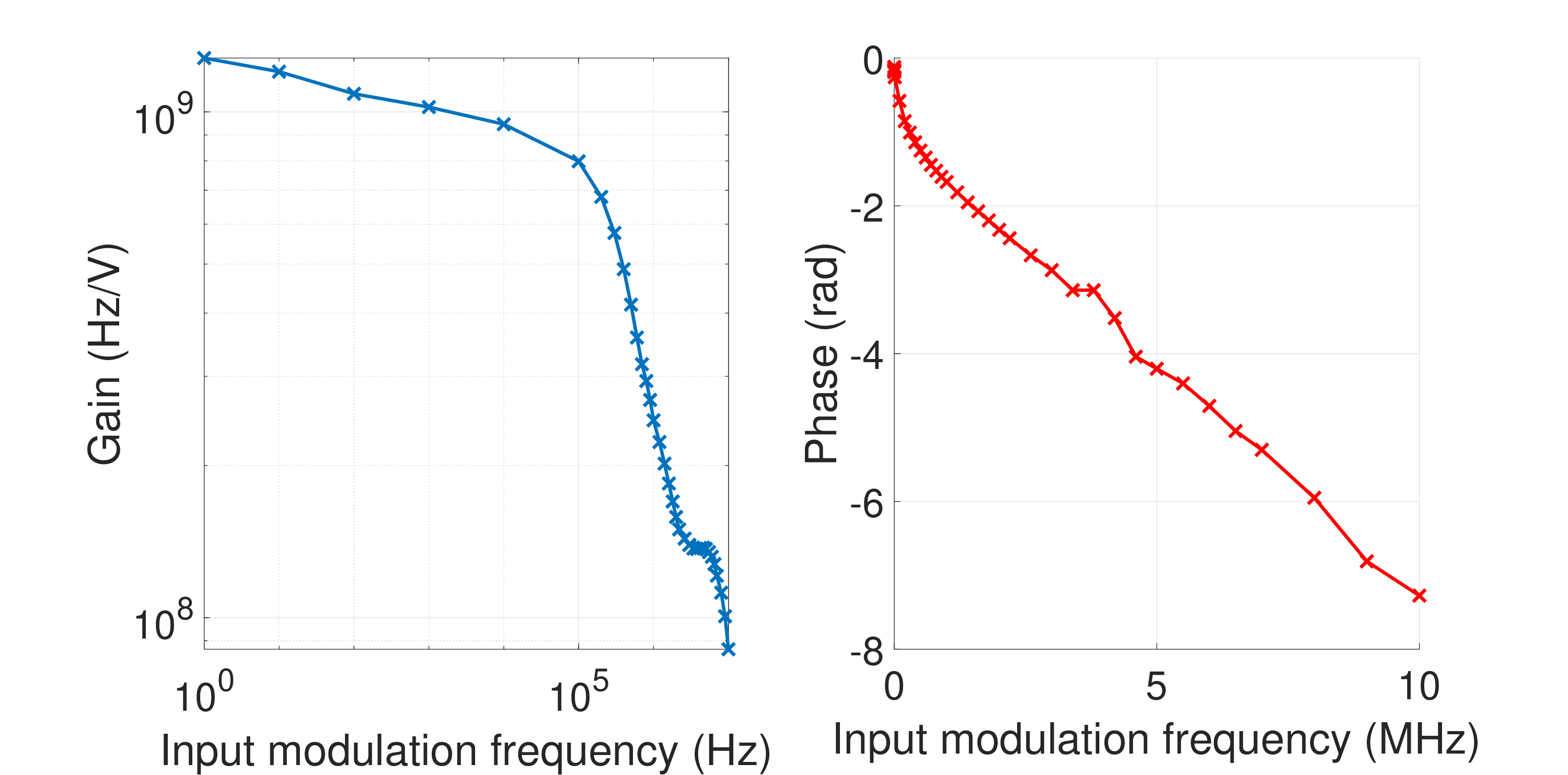}
\caption{Laser transfer function amplitude and phase when disturbed through the "Current Servo Input" modulation input of the Vescent D2-105 driver. The laser instantaneous frequency is  measured with the unbalanced, frequency-shifting MZI. Therefore this transfer function includes the response of this modulation input, that of the laser itself, and the response of the MZI.}
\label{fig:transferfunc}
\end{figure}

\begin{figure}[t]
 \centering \includegraphics[width=8.5cm]{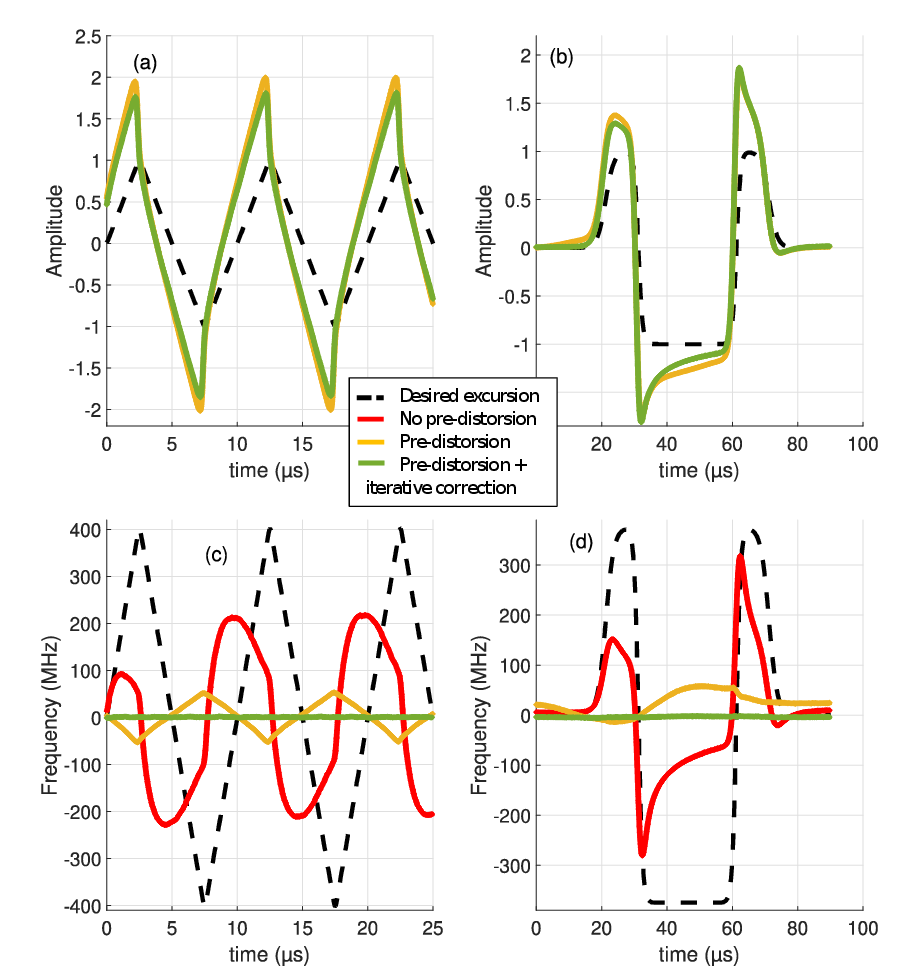}
\caption{Pre-distorted voltage commands (yellow line) and iteratively-corrected voltage commands (green line) in order to achieve the desired frequency excursions (black dashed line), in the case of (a) triangular and (b) arbitrary excursion. (c) and (d): frequency errors with a free-running laser (red line), with the pre-distorsion (yellow line) and after 20 iterations (green line).}
\label{fig:precorrection}
\end{figure}

In order to implement the pre-distorsion stage, we determine the laser transfer function $H(f)$ in amplitude and phase by sending sinusoidal voltages into the modulation input of the laser current controller and measuring the resulting instantaneous frequency variation of the laser (see Fig.~\ref{fig:transferfunc}). The voltage commands corresponding to triangular and arbitrary frequency excursions are then pre-distorted using  Eq.~\ref{eq:precorrection}. The pre-distortion, operating as an anticipation of the non-instantaneous laser response, leads to a significant improvement of the laser behaviour (see Figure~\ref{fig:precorrection}): the frequency errors are clearly reduced (of the order of $60$~MHz). For the triangular chirps, the RMS frequency error is reduced to $20.52$ MHz, equivalent to a $2.5\%$ relative RMS.

\begin{figure}[ht]
\centering \includegraphics[width=8.5cm]{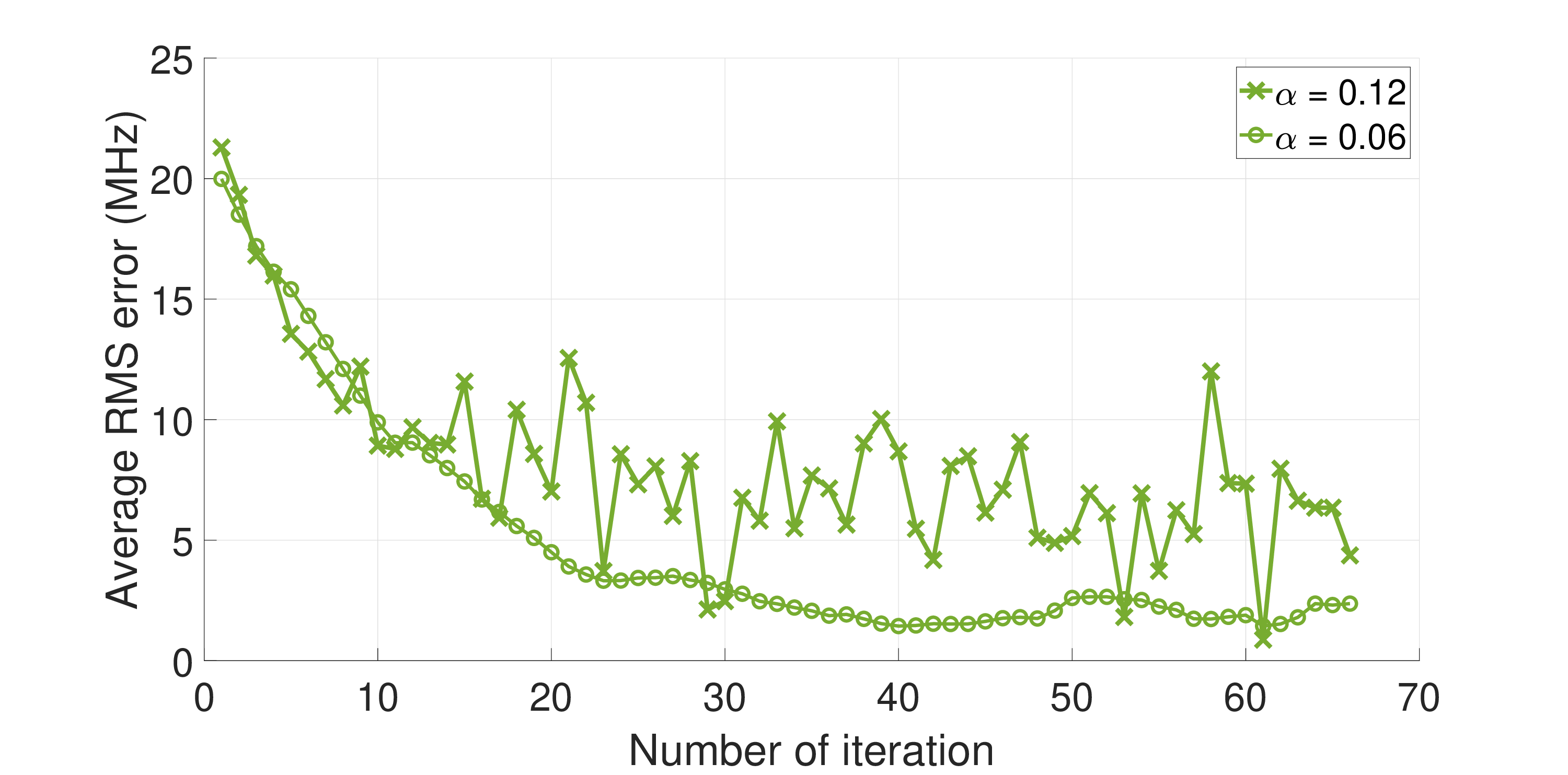}
\caption{Evolution of the RMS frequency error using the iteration algorithm described in Equation~\ref{eq:iteration}, in the case of a triangular $800$~MHz amplitude frequency excursion with $100$~kHz modulation frequency, for two values of the parameter $\alpha$.}
\label{fig:iterations}
\end{figure}

In order to further reduce the frequency error, we implement the iterative correction algorithm on both frequency excursions using Equation~\ref{eq:iteration} (see Figure~\ref{fig:precorrection}). We observe that after few tens of iterations, the frequency error is substantially reduced, down to a few MHz. A $2.1$ MHz RMS frequency error is obtained for the triangular chirps, corresponding to a $0.26\%$ relative RMS. Examining the convergence of the iterative algorithm, we show that larger $\alpha$ values allow a faster convergence but a lower frequency excursion precision (see Figure~\ref{fig:iterations}).

\section{Suppression of non-reproducible errors}
\label{sec:suppression}
Once the systematic errors are efficiently suppressed, one is left with only stochastic frequency errors, mostly coming from the intrinsic laser frequency noise, but also potentially caused by the steep frequency chirps and abrupt slope changes. We propose a versatile correction able to handle arbitrary frequency commands including fast laser chirps and monochromatic laser operation.

\subsection{Phase-Locked-Loop feedback correction}
\label{sec:feedback}
Self-heterodyne systems (such as the unbalanced MZI presented in Section~\ref{sec:mzi}) enable the generation of a signal proportional to the frequency offset via the downconversion of the detected beatnote with a local oscillator (LO). This signal can then be used as a correction signal to the laser source, resulting in a feedback loop. This setup is analog to an electronic phase-locked-loop (PLL) where the voltage-controlled oscillator (VCO) is composed of the laser source and the interferometer, and the frequency mixer and subsequent filter act as a phase comparator.

We use a versatile version of this setup that consists in using a specific local oscillator $V_{th}(t)$, proportional to the ideal beatnote signal that is expected for an ideal, error-free and noise-free laser frequency excursion, given any arbitrary laser frequency excursion~\cite{satyan2009precise}.

\begin{figure}[t]
\centering 
\includegraphics[width=8.5cm]{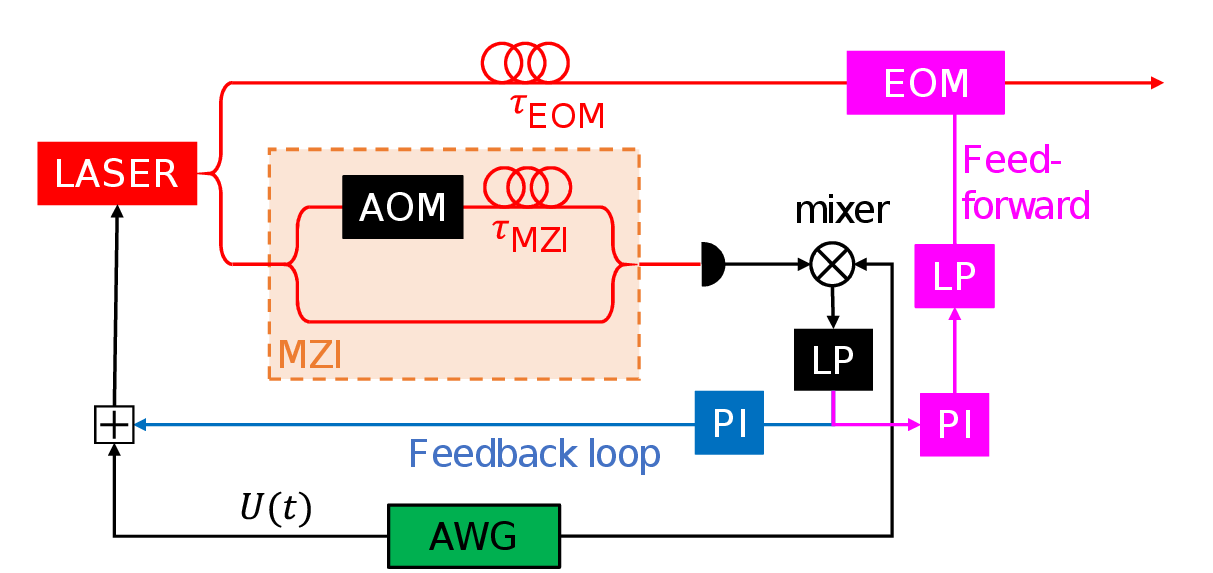}
\caption{Experimental setup including the feedback loop and the feed-forward correction for arbitrary frequency excursions. The laser current is driven via an arbitrary voltage command generated by an arbitrary waveform generator (AWG) fed into a fast modulation input port. For both corrections, an error signal is produced via the downconversion of the beatnote at the output of the self-heterodyne MZI. For the feedback loop, this error signal is filtered through a proportional-integrator filter (PI) and added to the voltage command to the laser current driver. For the feed-forward correction, the same error signal is filtered with a PI and a low pass filter (LP), and fed into a electro-optic phase modulator (EOM) after a physical fibered delay line on the user laser beam.}
\label{fig:setup}
\end{figure}

We experimentally implement this PLL feedback loop using the optical beatnote collected at the output of the MZI (see Section~\ref{sec:mzi}) and mixing it with the ideal beatnote signal generated by our multichannel AWG5004, using an electronic mixer (Mini-Circuits ZFM.1+). The use of a different channel from the same AWG for driving the AOM and creating the LO signal ensures minimal electronic jitter. The resulting signal is filtered with a $10$-MHz, $5$th order Tchebychev low-pass filter. A proportional-integrator loop filter is then used, built with a Newport LB1005S servo-controller operated in Low Frequency Gain Limitation (LFGL) mode with a $1$-MHz PI corner.

\begin{figure}[t]
\centering \includegraphics[width=8.5cm]{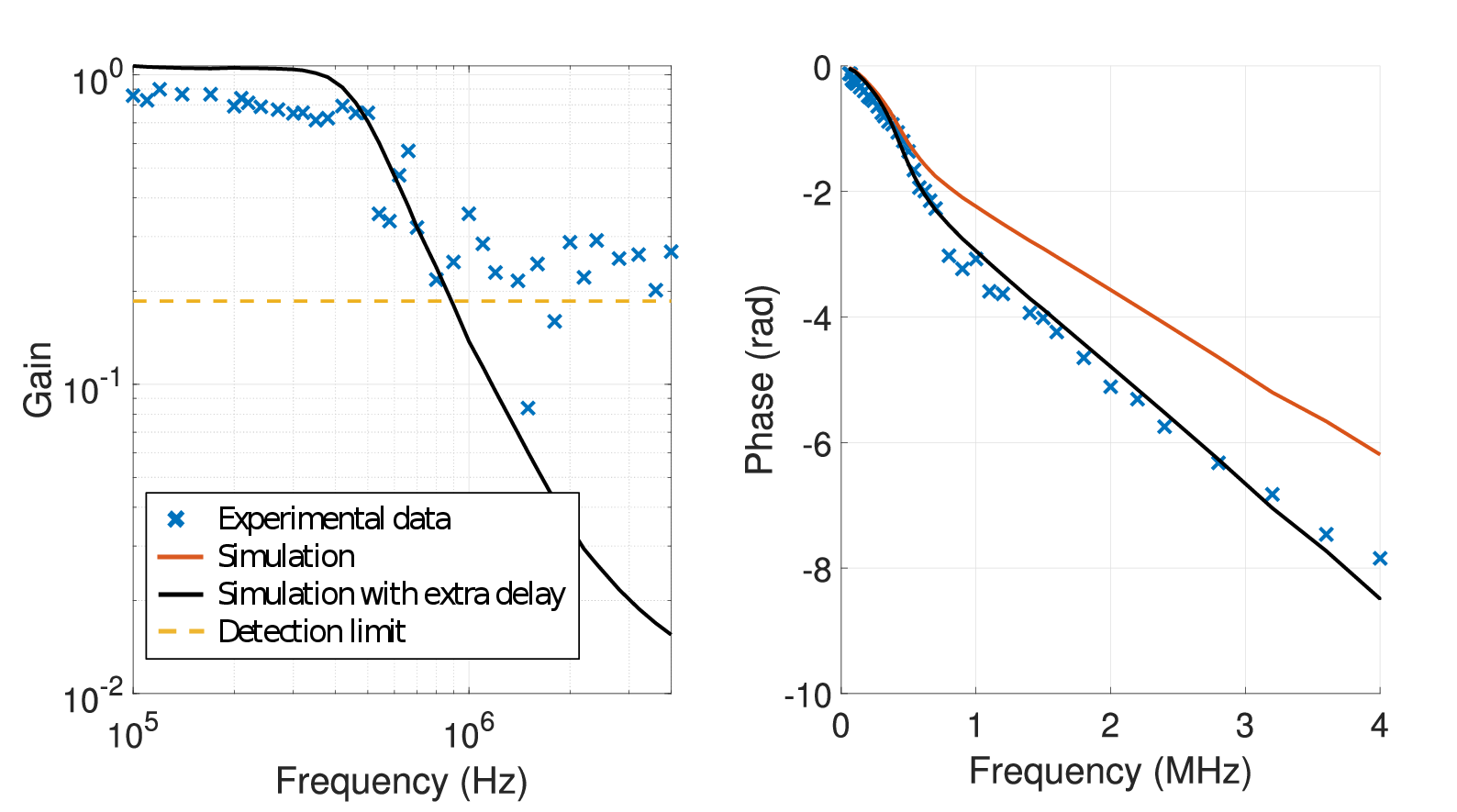}
\caption{Experimental (dots) and simulated (lines) closed feedback loop transfer function in amplitude (a) and phase (b). The simulated transfer function is derived from the independent measurement of each component of the loop (see Appendix~\ref{app:PLL_transfer}), completed with an additional delay (black lines). The red line shows the transfer function phase without this additional delay. The yellow dashed line shows the detection noise limit. }
\label{fig:plltransfer}
\end{figure}

The transfer function of the closed feedback loop is measured and plotted in Figure~\ref{fig:plltransfer}. A description of the measurement method is given in Appendix~\ref{app:feedback}. We observe a sharp cutoff around $600$~kHz. The PLL transfer function exhibits a quasi-linear phase, whose slope corresponds to an effective group delay $\tPLL=270$~ns where $\Phi(f) = 2\pi f\tPLL$. We demonstrate in Appendix~\ref{app:feedback} that this delay corresponds to the physical delay of the various optical and electronic elements in the loop, and that it directly affects the feedback bandwidth via $BW=\frac{1}{8 \tau_{PLL}} \simeq 460$~kHz.

\subsection{Feed-forward correction}
\label{sec:ff}
The high frequency fluctuations triggered by fast frequency excursions require an instantaneous frequency correction with a broad bandwidth. To overcome the feedback bandwidth limitation, we propose to combine it with a feed-forward (FF) correction. Feed-forward consists in applying a correction to the laser emission a posteriori, using the knowledge of the current laser behaviour. To achieve this, a phase-modulating device must be used, such as an electro-optic modulator (EOM) or acousto-optic modulator (AOM). An EOM offers a weaker tunability than an AOM but a higher bandwidth, which is key for the correction of high frequency errors. The FF approach has already been proposed for laser linewidth narrowing, either independently~\cite{aflatouni2012wideband} or in combination with feedback~\cite{cheng2022feed,lintz2017note}.

In our design, we propose to generate a FF correction based on the same error signal as the feedback loop after the downconversion step. This error signal is sent to the phase modulator inserted in the user beam (see Fig.~\ref{fig:setup}), after appropriate electronic filtering. Importantly, the quality of the FF correction relies on the application of the correction at the right time. Therefore, to account for the delay naturally occurring in the buildup of the correction signal, a physical delay is inserted before the phase modulator along the user beam.

We implement the FF scheme using a fibered electro-optic fibered phase modulator (iXblue MPXLN-0.3) with $300$~MHz bandwidth. The FF correction signal is obtained using the same error signal as the PLL fed into a specific feed-forward filter composed of another Newport LB1005S servo-controller operated in LFGL mode ($1$MHz PI-corner) and a low-pass RC filter with $10$~MHz cutoff frequency. The combination of all the optical and electronic elements of the feed-forward correction yields a high-pass behaviour and a total group delay of around $180$~ns (see Appendix~\ref{app:feedforward}). In order for the FF correction to reach the laser emission at the right time, a $28$~m fibered delay line (corresponding to a $140$~ns delay) is inserted before the EOM.

We investigate the influence of this physical delay on the quality of the noise compensation and show that it has a critical effect on the quality of the feed-forward correction and in turn on the laser linewidth (see Appendix~\ref{app:feedforward}). Interestingly, in contrast with the feedback correction, the feed-forward bandwidth is only limited by the bandwidth of the various elements of the FF correction signal and remains independent from the feed-forward delay as long as the latter is perfectly matched by the physical delay.

\subsection{Implementation on a monochromatic laser}
\begin{figure}[t]
\centering \includegraphics[width=8.5cm]{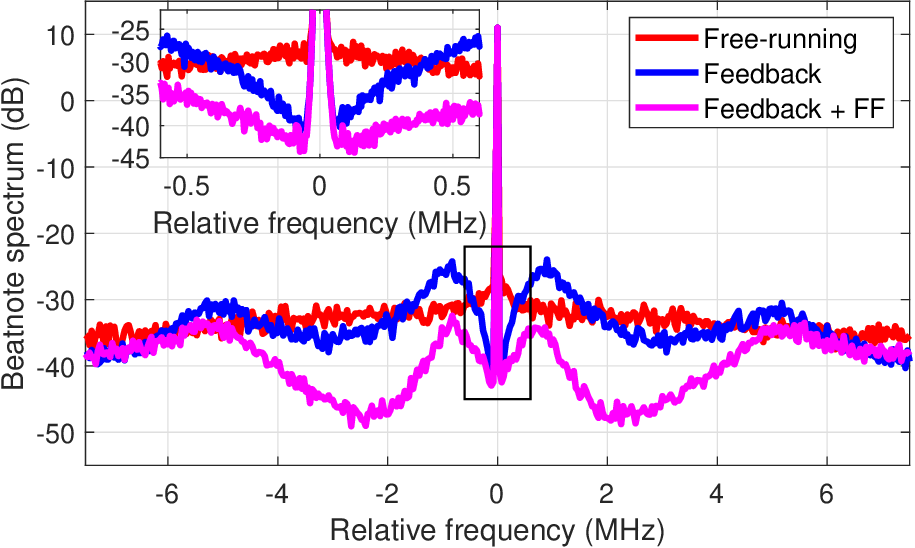}
\caption{Self-heterodyne beatnote spectra with a monochromatic laser (resolution bandwidth: $10$ kHz) for various experimental configurations. The beatnote spectrum is centered on the $80$~MHz driving frequency of the AOM inside the MZI (see Fig.~\ref{fig:mzi}). Inset: close-up view showing the feedback bandwidth. }
\label{fig:psd_fb_ff}
\end{figure}

Our dynamic, real-time corrections are tested on the DBR laser in monochromatic operation. In Figure~\ref{fig:psd_fb_ff} we plot the power spectral density (PSD) of the beatnote signal with and without feedback correction. We obtain a noise rejection of about $10$~dB in a $150$~kHz range around the carrier frequency. One can derive the feedback bandwidth as the frequency at which the beatnote signal PSD reaches the value of the free-running laser beatnote PSD. Such an estimation yields a feedback bandwidth around $400$~kHz, in agreement with the bandwidth derived from the closed loop transfer function (see Section~\ref{sec:feedback}).

\begin{figure}[t]
\centering
\includegraphics[width=7.5cm]{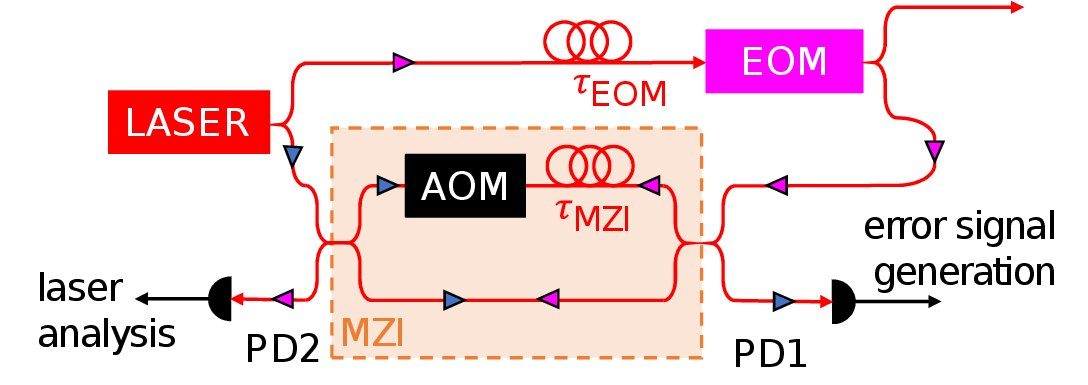}
\caption{Simultaneous error signal generation and instantaneous laser frequency measurement with a single MZI: the MZI is assembled with $2\times2$ 3dB fiber couplers allowing two opposite, independent propagation schemes in the same equipment and ensuring perfect relative stability.}
\label{fig:backwards}
\end{figure}

Assessing the influence of the FF correction on the laser spectral purity is less straightforward. Indeed, a second optical frequency discriminator (OFD) is needed~\cite{lintz2017note}. However, using a separate, independent device would lead to relative instability due to uncorrelated acoustic and thermal perturbations. To avoid this we use the \emph{same} physical OFD as the one used for the error signal generation (see Fig.~\ref{fig:backwards}): we simultaneously send the feed-forward-corrected laser beam in the unused port of the MZI, in the backwards direction. The effect of the full correction can then be analyzed via the spectrum of the beatnote collected on the second photodiode (see Figure~\ref{fig:psd_fb_ff}). We demonstrate a $5$~MHz correction bandwidth for the combined feedback and feed-forward stages. The maximum noise rejection (15dB) occurs at $2.4$~MHz from the carrier frequency.

\begin{figure}[t]
\centering
\includegraphics[width=8.5cm]{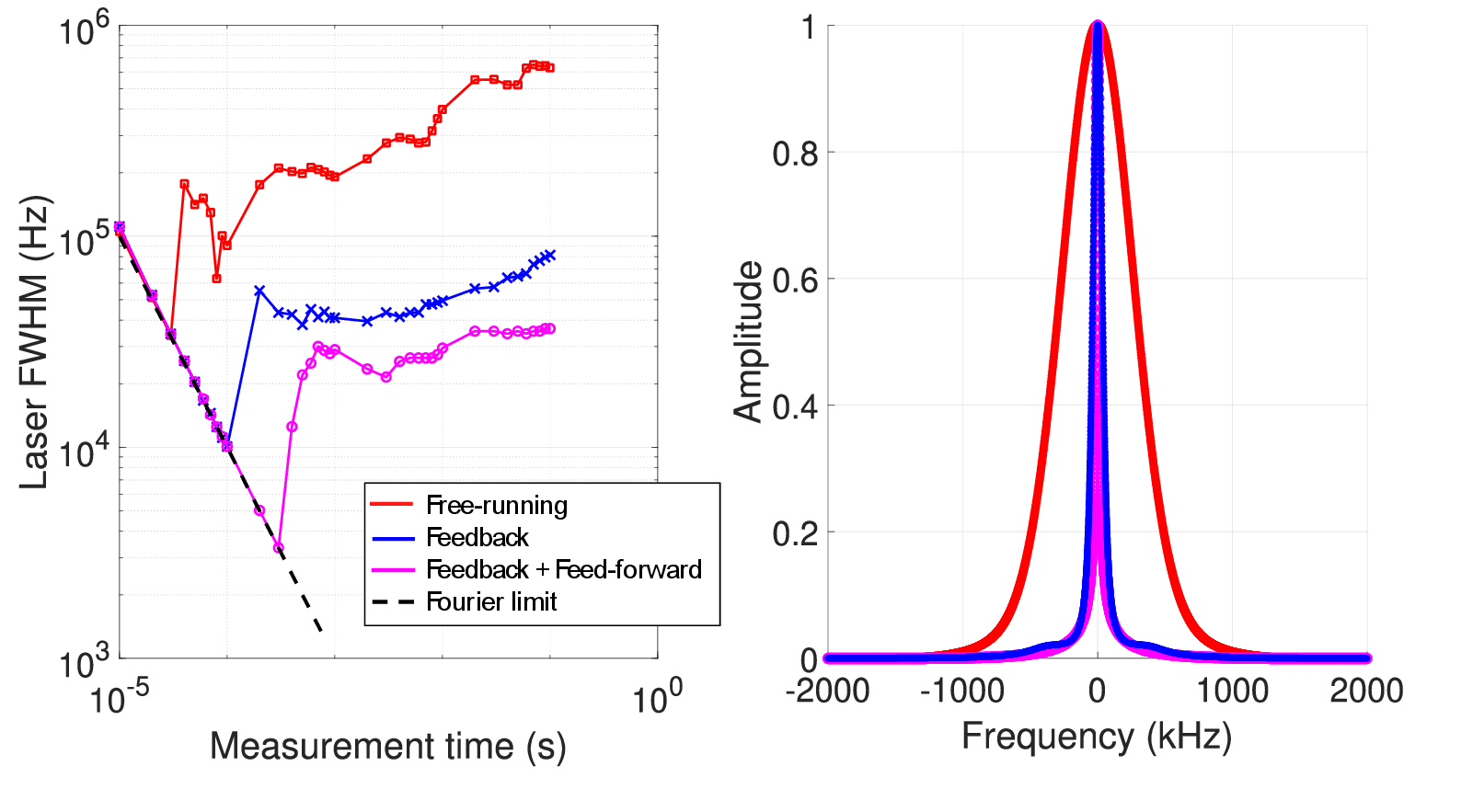}
\caption{Reconstruction of the laser spectrum in monochromatic operation using the calibrated output signal of a commercial optical frequency discriminator. (a) Laser linewidth (estimated as the FWHM of the reconstructed laser lineshape) for various integration times between $10~\mu$s and $100$~ms. The dashed black line represents the Fourier uncertainty that limits the linewidth derivation~\cite{vonbandel2016time}. (b) Reconstructed laser lineshape for free-running laser (red), feedback-locked laser (blue) and feedback + feed-forward locked laser (magenta), where the frequency noise is integrated over $100$ ms.}
\label{fig:laser_linewidth}
\end{figure}

We can also analyze the performance of our feedback and feed-forward corrections by examining the laser linewidth and laser spectrum that we derive from the output of a commercial optical frequency discriminator (Silentsys OFD)~\cite{didomenico2010simple} (see Figure~\ref{fig:laser_linewidth}). We observe the typical linewidth tendency to increase with the integration time~\cite{vonbandel2016time}, in all configurations. The feedback has a crucial impact on the linewidth on the investigated timescales (above $10~\mu$s). The feed-forward correction, on the other hand, is most efficient on shorter timescales where the linewidth cannot be accurately measured. The $740$-kHz linewidth of the free-running laser for a $100$~ms integration time is reduced down to $81$~kHz with the feedback correction, and further down to $30$~kHz with both the feedback and the feed-forward corrections. This value is only marginally larger than the laser's $10$~kHz Lorentzian linewidth limit attributed to the white part of its residual frequency noise~\cite{stephan2005laser}.

\section{Implementation of the four-stage correction on complex laser chirps}

In this section we test our complete, multistage correction (including pre-distorsion, iterative correction, feedback and feed-forward) on complex  frequency excursions on our DBR laser. Two configurations are studied, namely triangular excursions, and the arbitrary excursion with $750$~MHz by $2.5~\mu$s chirps described in Section~\ref{sec:exp1}. 

It is important to realize that the feedback and feed-forward correction can only correct relatively small frequency errors. The initial pre-distorsion and iterative corrections, limiting the error to within a few MHz, are an absolute pre-requisite to ensure the proper action of the real-time correction stages. In their absence, the laser frequency error may largely exceed the MZI free spectral range, making the error signal unreliable.

\subsection{Triangular excursions}
\begin{figure}[t]
\centering \includegraphics[width=8.5cm]{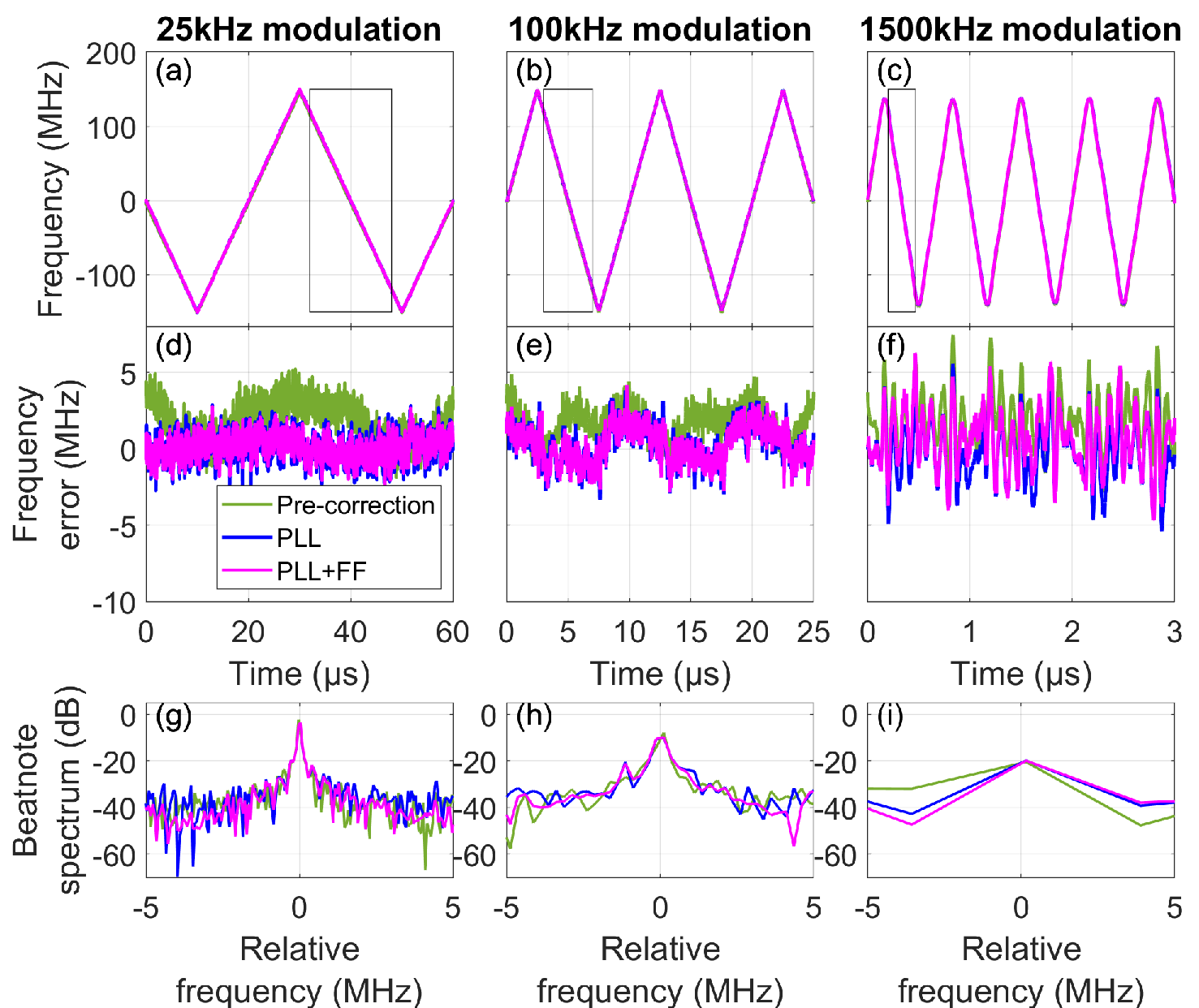}
\caption{Implementation of the feedback and feed-forward correction stages on $300$~MHz peak-peak triangular excursions with $25$, $100$ and $1500$ kHz modulation frequencies. The green lines correspond to the free-running laser behaviour, after pre-distorsion and iterative correction of the command voltage. The blue and magenta lines correspond to the laser behaviour with the combination of pre-distortion and iterative correction with feedback correction (blue), and with feedback and feed-forward correction (magenta). For every configuration we plot the frequency excursions (a, b, c), the frequency errors (d, e, f), and the beatnote spectra (g, h, i). The latter are computed on the rectangles shown in (a) (b) and (c), corresponding to 90\% of the downwards linear chirp. For fast modulation rates (i), the beatnote spectrum is computed over a sub-$\mu$s timescale, making the resolution bandwidth very low.}
\label{fig:triangle_PLL_FF}
\end{figure}

The feedback and feed-forward corrections are implemented on triangular frequency excursions, using the pre-distorted and iteratively corrected voltage command obtained as described in Section~\ref{sec:systematic}. The laser instantaneous frequency is measured using the backwards MZI (see Figure~\ref{fig:backwards}) and compared to the desired frequency excursion. The instantaneous frequency error and the optical beatnote spectrum are shown in Figure~\ref{fig:triangle_PLL_FF}. The corresponding RMS frequency error,  computed on several periods including the abrupt slope changes, is given in Table~\ref{tab:RMS_triangles}.

\begin{table}
\centering
\small
\begin{tabular}{cccccc}
 \hline
 \hline
 $f_\mathrm{modulation}$& $\sigma_\mathrm{iteration}$ & $\sPLL$& $\sigma_\mathrm{PLL+FF}$ & relative RMS\\
 \hline
 $25$ kHz   & $2.17$~MHz & $0.79$~MHz & $0.68$~MHz & $0.23\%$\\
 $100$ kHz  & $1.99$~MHz & $1.24$~MHz & $1.19$~MHz & $0.40\%$\\
 $1500$ kHz & $2.77$~MHz & $2.19$~MHz & $2.06$~MHz & $0.69\%$\\
 \hline
 \hline
\end{tabular}
\caption{Frequency fluctuation performances for triangular excursions with $300$ MHz span. Note that the value of the RMS is computed over several periods of the triangular wave, including the abrupt slope changes.}
\label{tab:RMS_triangles}
\end{table}

The feedback correction visibly enhances the overall chirp precision for all the configurations shown, notably in terms of RMS frequency error. Nevertheless, the feedback bandwidth limits the correction efficiency as one reaches for higher modulation frequencies.

The feed-forward correction further improves the frequency excursion by correcting fluctuations at a faster rate. This is particularly visible in the beatnote spectra where the feed-forward significantly reduces the noise a few MHz away from the center frequency.
Nevertheless the sharp angles of the triangular command lead to residual high frequency fluctuations, even with the feed-forward part (see Table~\ref{tab:RMS_triangles}). We attribute this limitation to the detection noise that is imparted onto the laser emission via the feedback and feed-forward (see Appendix~\ref{app:feedback}). The relative RMS remains well below $1\%$ for all the configurations shown. Note that due to its low-frequency cutoff, the feed-forward correction cannot operate without the feedback correction, due to significant slow frequency noise.

In Table~\ref{tab:Comparison} we provide a brief review of recent publications addressing steep laser linear chirps. We compare them with the full four-stage correction presented in Fig.~\ref{fig:triangle_PLL_FF}, using the RMS frequency error computed only in the linear sections of the triangular excursions, \emph{ie} avoiding the slope changes. The MHz-range precision that we reach, even with extreme chirp rates, demonstrates the versatility of our approach.

\begin{table}
\centering
\small
\begin{tabular}{cccccc}
\hline
\hline
Reference & Frequency span & Sweep duration & Chirp rate & RMS frequency error in linear section \\
\hline
\cite{qin2015coherence}           & 50 GHz & $100$~ms & 0.5 THz/s & 0.09~MHz \\
\cite{roos2009ultrabroadband}   & $4.8$~THz & $0.8$~s & 6 THz/s & $0.2$ MHz\\
This work                       & 300 MHz & $20~\mu$s & 15 THz/s & 0.74 MHz\\
\cite{lihachev2023frequency}    & 1.8 GHz & 50 $\mu$s & 36 THz/s & 0.9 MHz \\
\cite{cao2021highly}                & 26 GHz & 0.5 ms & 52~THz/s & 1.5 MHz \\
This work                       & 300 MHz & $5~\mu$s & 60 THz/s & 0.83 MHz\\
\cite{kervella2014laser}        & 20 GHz & $250~\mu$s & 80 THz/s & $<1$ MHz \\
\cite{zhang2019laser}           & 36 GHz & 100 $\mu$s & 360 THz/s & 1 MHz\\
\cite{li2022linear}              & 25 GHz & 50 $\mu$s & 500 THz/s & 1.04 MHz \\
This work                    & 300 MHz & 0.33 $\mu$s & 910 THz/s & 1.63 MHz \\
\hline
\hline
\end{tabular}
\caption{Comparison of chirp linearities with recently published works, ordered by increasing chirp rates.}
\label{tab:Comparison}
\end{table}

We also test our full multi-stage correction on triangular shapes with increasing chirp span $\Delta f$, keeping the modulation frequency at $25$ kHz and $100$ kHz. The results are presented in Table~\ref{tab:RMS_triangles4}.
Consistently with our earlier findings, we find that each correction stage successfully improves the frequency excursion accuracy. 
Interestingly, we show that the relative RMS improves with increasing chirp span, reaching approximately $0.1\%$ with GHz chirp spans.

\begin{table}
\centering
\small
\begin{tabular}{cccccc}
 \hline
 \hline
 $f_{mod}$ & $\Delta f$ & $\sigit$ & $\sPLL$& $\sigma_{PLL+FF}$ & relative RMS\\
 \hline
 \multirow{3}{*}{$100$kHz} & $300$MHz &  $1.99$MHz & $1.24$MHz & $1.19$~MHz & $0.40\%$\\
 &$550$MHz & $1.94$MHz & $1.43$MHz & $1.14$MHz & $0.21\%$\\
 &$800$MHz & $2.26$MHz & $1.55$MHz & $1.46$MHz & $0.18\%$\\
 \hline
 \multirow{3}{*}{$25$kHz} & $300$MHz & $2.17$MHz & $0.79$MHz & $0.68$MHz & $0.23\%$\\
 & $750$ MHz & $2.48$MHz & $1.19$MHz & $1.04$MHz & $0.14\%$\\
 & $1200$ MHz & $2.26$MHz & $1.55$MHz & $1.46$MHz & $0.12\%$\\
 \hline
 \hline
\end{tabular}
\caption{Frequency fluctuation RMS frequency error and relative RMS for triangular excursions with various modulation frequencies and chirps spans.}
\label{tab:RMS_triangles4}
\end{table}

\subsection{Arbitrary excursions}
\begin{figure}[t]
\centering \includegraphics[width=8.5cm]{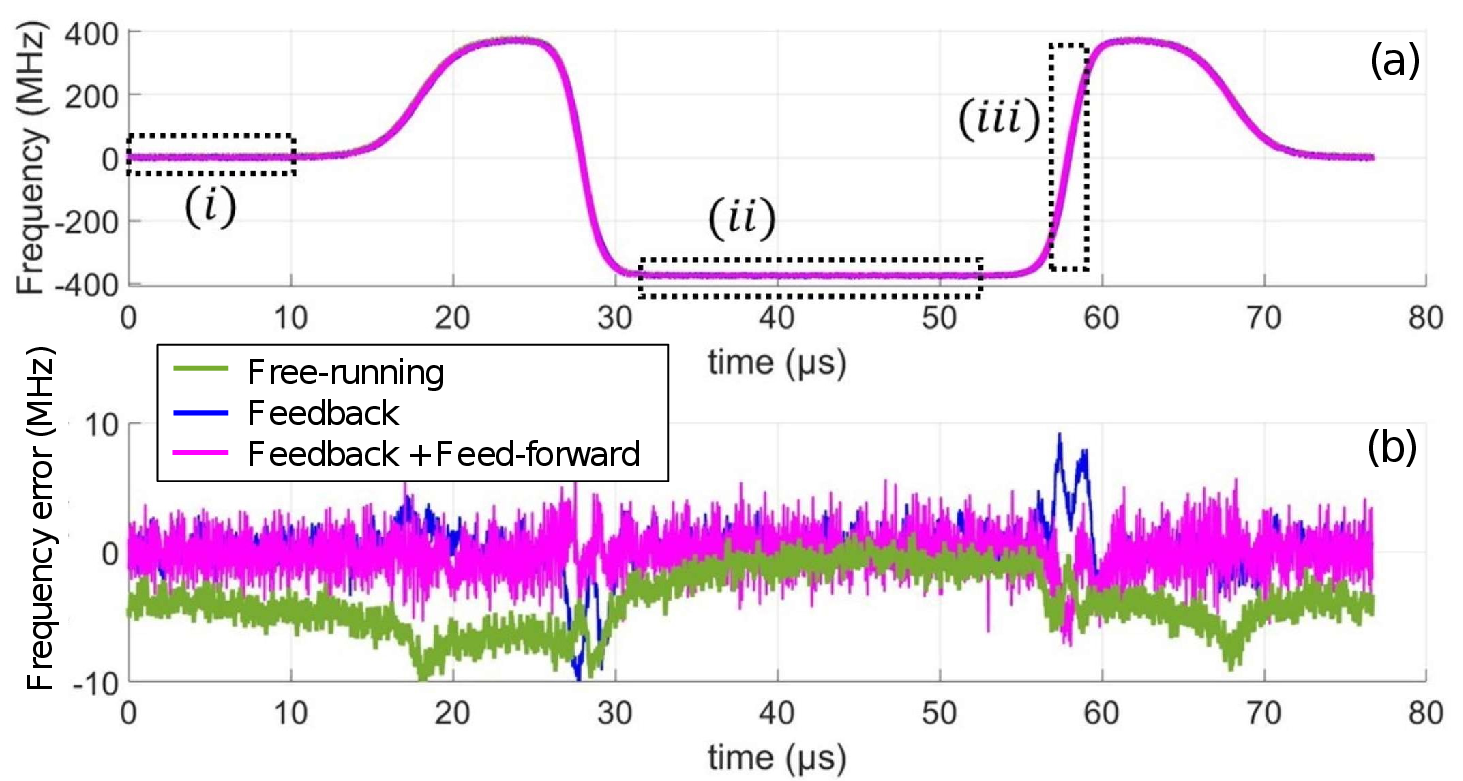}
\caption{Laser response (a) and respective frequency errors (b) for an arbitrary frequency excursion. The green line corresponds to the free-running laser behaviour using the pre-distorted and iteratively corrected voltage command. The blue and magenta lines correspond to the same voltage command combined with feedback correction, and combined with feedback correction and FF correction, respectively. The RMS frequency error is retrieved in the three rectangular frames: $\sigma_\mathrm{FF}=1.45$~MHz, $1.62$~MHz and $1.42$~MHz for $(i)$, $(ii)$ and $(iii)$, respectively.}
\label{fig:arb_PLL_FF}
\end{figure}

We now study the effect of our multi-stage correction on the arbitrary frequency excursion, using the pre-distorted and iteratively corrected voltage command described in Section~\ref{sec:systematic} (see Fig.~\ref{fig:arb_PLL_FF}).

Similarly to the triangular excursions, the feedback loop only corrects the low frequency fluctuations and adds high-frequency noise. The feed-forward correction efficiently complements the feedback loop as it strongly reduces the remaining frequency excursion imprecisions that occurred at the steep frequency variation points. Remarkably, the RMS frequency error evaluated on the intermediate monochromatic stage $(ii)$ is only marginally deteriorated with respect to the initial RMS error before the $800$~MHz-range perturbation $(i)$. We compute the laser lineshape in the $22~\mu$s interval between the two chirps $(ii)$ and obtain a $48$~kHz linewidth, close to the Fourier-transform limit, demonstrating the excellent laser spectral purity although in the midst of an abrupt frequency excursion. Finally, the RMS error in the linear part of the positive slope ($500$~MHz/$2~\mu$s) at the end of the arbitrary sequence $(iii)$ is as low as that in monochromatic operation $(i)$. This highlights our multi-stage correction's versatility, \emph{ie} its ability to operate on a succession of abrupt frequency chirps and monochromatic operation.

\section {Conclusion}
A current-modulated laser can be used to generate fast arbitrary frequency excursions with high modulation frequency and large chirp span. However, the inaccuracy of the frequency excursion and the high-frequency fluctuations induced by the fast spectral components of the modulation calls for an efficient and broadband correction design. In this work, a multi-stage correction is developed, including pre-distorsion and iterative correction to suppress systematic errors, and the combination of a feedback loop and a feed-forward correction both based on a self-heterodyne, unbalanced interferometer to suppress both slow and fast stochastic errors up to $5$~MHz.

We test our multi-stage correction on various laser frequency excursions to demonstrate its efficiency and versatility. In monochromatic operation we demonstrate an impressive reduction of the laser linewidth with the combined feedback and feed-forward stages. We also achieve $300$-MHz triangular excursions with a relative RMS below $1\%$ with modulation frequencies up to $1500$~kHz. Finally, we implement this multi-stage correction on an arbitrary frequency excursion with alternating steep slopes and monochromatic segments, and consistently demonstrate $<100$kHz laser spectral purity even in the midst of GHz-scale excursions.

This work opens new possibilities in various application domains including FMCW lidar, telecommunications or wideband signal processing, where fast and accurate laser frequency excursions are required.



\appendix
\section{Appendix: Parameters of the feedback loop}
\label{app:feedback}
In this section we discuss the relation between the feedback loop parameters and its bandwidth. First, PLL general considerations are presented and applied to our MZI-based feedback loop. Then the influence of the PLL delay on the feedback bandwidth and error correction is studied. Finally, the MZI delay impact on the laser instantaneous frequency measurement and the buildup of the error signal is discussed.

\subsection{PLL transfer function}
\label{app:PLL_transfer}


A PLL generally contains the following elements: a local oscillator, a phase comparator, a loop filter, and a voltage-controlled ocillator (VCO)~\cite{blanchard1976pll}. The Open Loop Transfer Function (OLTF) is the product of the transfer functions of the successive elements of the loop:
\begin{equation}
    F_\mathrm{OL}(f) = \frac{1}{2i\pi f}K_{\Phi} F_\mathrm{lf}(f) K_\mathrm{VCO}(f),
    \label{eq:oltf}
\end{equation}
where $K_{\Phi}$ is the phase comparator gain such that, when two signals with a $\Delta\Phi$ phase difference are sent to the phase comparator, the following signal is output: $V = K_{\Phi} \Delta \Phi$. $F_\mathrm{lf}(f)$ is the loop filter transfer function, and $K_\mathrm{VCO}(f)$ is the VCO transfer function. Finally, the Closed Loop Transfer Function (CLTF) is given by:
\begin{equation}
F_\mathrm{CL}(f) = \frac{F_\mathrm{OL}(f)}{1+F_\mathrm{OL}(f)}
\label{eq:CLTF}
\end{equation}

In the case of feedback correction on a current-controlled laser (see Section~\ref{sec:suppression} and Figure~\ref{fig:setup}), the VCO includes the laser driver, the laser source and the MZI. Its transfer function, given by $H(f)$, is described and measured in section~\ref{sec:exp1} (see Fig.~\ref{fig:transferfunc}). The loop filter is comprised of the servo-controller in PI mode, and a passive, 5-th order low-pass filter with a $10$~MHz cutoff frequency, whose transfer functions are measured separately. The phase comparator, whose role is ensured by the mixer, is also characterized and found to give rise to a significant group delay together with a constant dephasing, so we write $F_\Phi(f)=K_\Phi e^{-2i\pi f\tmix-i\Phi_\mathrm{mix}}$, with $\tmix=50$~ns and $\Phi_\mathrm{mix}=\pi/4$. Finally the various delays induced by each element of the feedback loop are summarized in Table~\ref{tab:fb_group_delays}. Finally the OLTF reads as:
\begin{equation}
    F_\mathrm{OL}(f) = \frac{1}{2i\pi f}  F_{\Phi}(f) F_{PI}(f) F_\mathrm{lp}(f) H(f)
    \label{eq:oltf_mzi}
\end{equation}

\begin{table}
\centering
\begin{tabular}{ccc}
\hline \hline
Component & Transfer function & Group delay\\
\hline
Laser+MZI & $H(f)$ & $100$~ns \\
LB1005 & $F_\mathrm{PI}(f)$ & $40$~ns \\
Mixer & $F_\Phi(f)$ & $50$~ns \\
Low-pass filter & $F_\mathrm{lp}(f)$ & $44$~ns \\
\hline \hline
 \end{tabular}
\caption{Effective group delays induced by each element of the feedback loop.}
\label{tab:fb_group_delays}
\end{table}

We measure the feedback loop transfer function $F_\mathrm{CL}(f)$ by making the laser frequency oscillate with the feedback loop closed. A small amplitude ($8$ MHz), monochromatic frequency modulation is applied on the local oscillator. The closed PLL ensures that the phase of the MZI output matches the phase of the local oscillator, which induces a frequency modulation of the laser source. The resulting instantaneous frequency variations of the laser emission are measured by analyzing the beatnote at the output of the MZI, giving access to the PLL transfer function plotted in Fig.~\ref{fig:plltransfer}. The dynamic range of the gain measurement is limited because the feedback loop parameters must be kept constant for all frequencies, rendering impossible to compensate for the strong attenuation of frequencies above $500$~kHz.
In order to accurately account for the experimental transfer function measurements, the simulated transfer function given by Eqs.~\ref{eq:CLTF} and \ref{eq:oltf_mzi} is corrected by a small additional delay ($\tau_\mathrm{extra}=101$~ns), ascribed to the various electronic and optical delays present in the loop and not explicitly considered in our analysis (free space propagation, coaxial cables, splitters, etc). The simulation finally provides a good understanding of the PLL transfer function, and allows us to estimate the total group delay associated to the feedback loop $\tPLL=335$~ns, accounting for all its components.

\begin{figure}[t]
\centering \includegraphics[width=8.5cm]{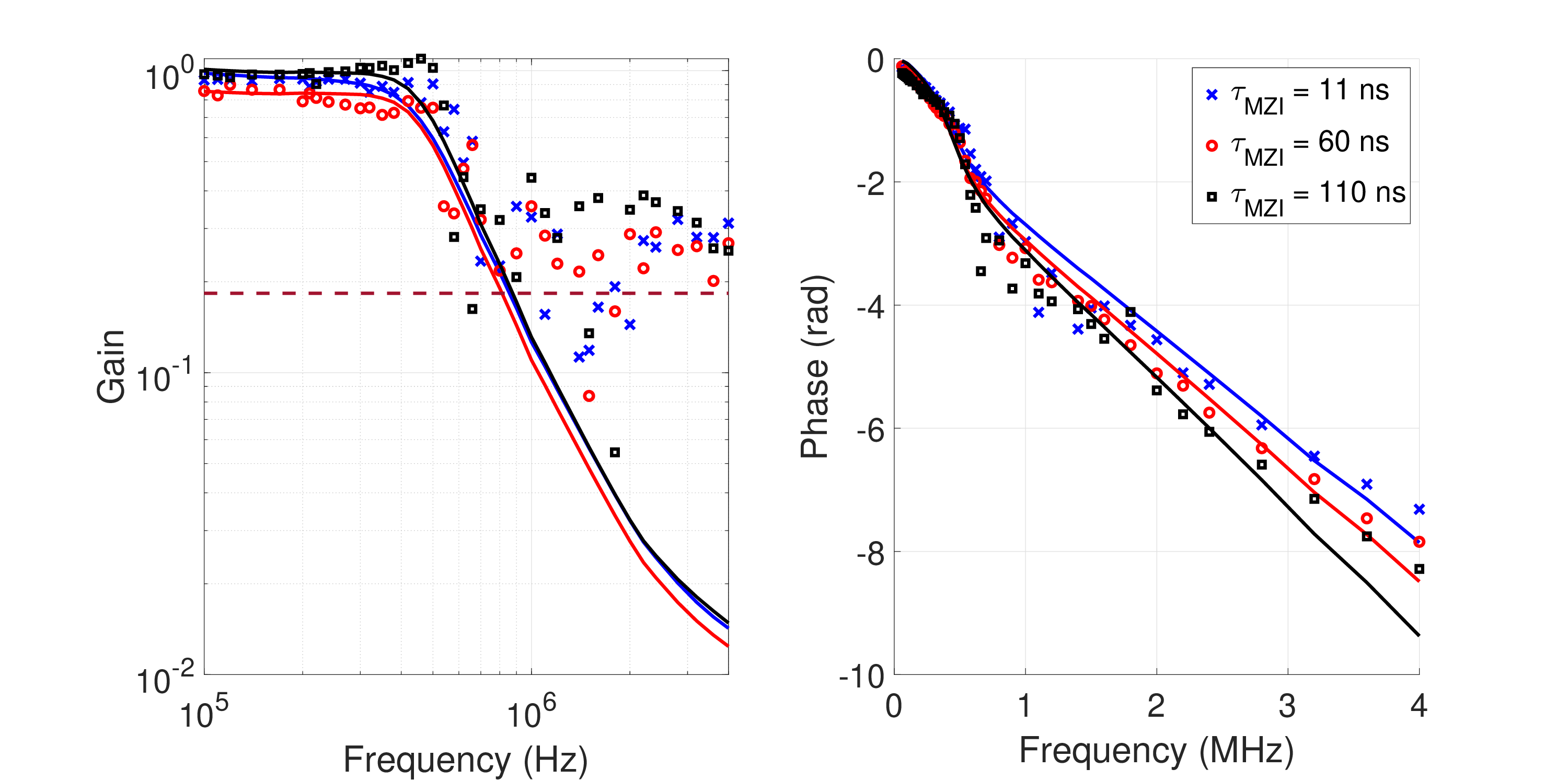}
\caption{PLL transfer function for 3 different MZI delays: $11$ ns, $60$ ns and $110$ ns. PLL transfer function measurements are plotted for 3 different MZI delays: $11$ ns (blue points), $60$ ns (red points) and $110$ ns (black points). The corresponding theoretical transfer functions for each MZI delay are computed using equations~\ref{eq:oltf} and~\ref{eq:CLTF} with the transfer functions of the PLL elements and the corresponding MZI delay: $11$ ns (blue line), $60$ ns (red line) and $110$ ns (black line)}
\label{fig:diffdelays}
\end{figure}

We now explore the impact of the MZI delay on the PLL transfer function. To that end we operate the laser with the feedback correction using various fiber lengths in the MZI (such that $\tMZI=11$~ns, $60$~ns and $110$~ns). The results plotted in Figure~\ref{fig:diffdelays} show only a weak dependence of the PLL transfer function, confirming that the MZI delay only contributes marginally to the feedback bandwidth.


\subsection{PLL delay and feedback bandwidth}
\label{app:plldelay}
In this section we examine how the group delay-like behavior of the PLL transfer function leads to a bandwidth limitation. Let us assume that the laser exhibits a sinusoidal frequency fluctuation $\epsilon_{0} (t)$ at frequency $F$ with an amplitude $\Delta F$: $
\epsilon_{0} (t) = \Delta F \sin(2 \pi F t)$.
If the feedback loop is equivalent to a mere delay line, the feedback signal reads as $U_{fb}(t) = G \epsilon_0(t-\tPLL)$, where $G$ is the feedback loop gain. After applying the feedback signal to the laser source, the frequency fluctuation becomes $\epsilon (t) = \epsilon_{0} (t) - U_{fb}(t)$.
To quantify the effect of this feedback on the laser frequency stability, we examine the variance $\sigma^{2}=\langle \epsilon(t)^2 \rangle$ of the frequency fluctuations:
\begin{equation}
    \sigma^2 = \Delta F^2 \left( \frac{G^2+1}{2} - G \cos (2\pi F \tPLL) \right)
    \label{eq:var_PLL}
\end{equation}
$\sigma^{2}$ reaches a minimum when $G=\cos(2\pi F\tPLL)$:
\begin{equation}
\sigma_{min}^2=\frac{\Delta F^2}{2} \sin^2(2\pi F\tPLL)
\end{equation}
The fluctuation variance with no feedback  is $\sigma_0^{2}=\frac{\Delta F^2}{2}$. The ratio $\frac{\sigma^{2}_{min}}{\sigma_0^{2}}$ can be seen as the feedback loop's ability to correct fluctuations at frequency $F$. To lowest order, this figure of merit quadratically increases with the fluctuation frequency $F$. We define the feedback loop bandwidth as the frequency $F$ for which the aforementioned variance ratio is equal to $\frac{1}{2}$:
\begin{equation}
     F_{BW}=\frac{1}{8\tPLL}
     \label{eq:pll_bw}
\end{equation}
This confirms the common knowledge that a large feedback bandwidth requires a short loop delay $\tPLL$.

\begin{figure}[t]
\centering \includegraphics[width=8.5cm]{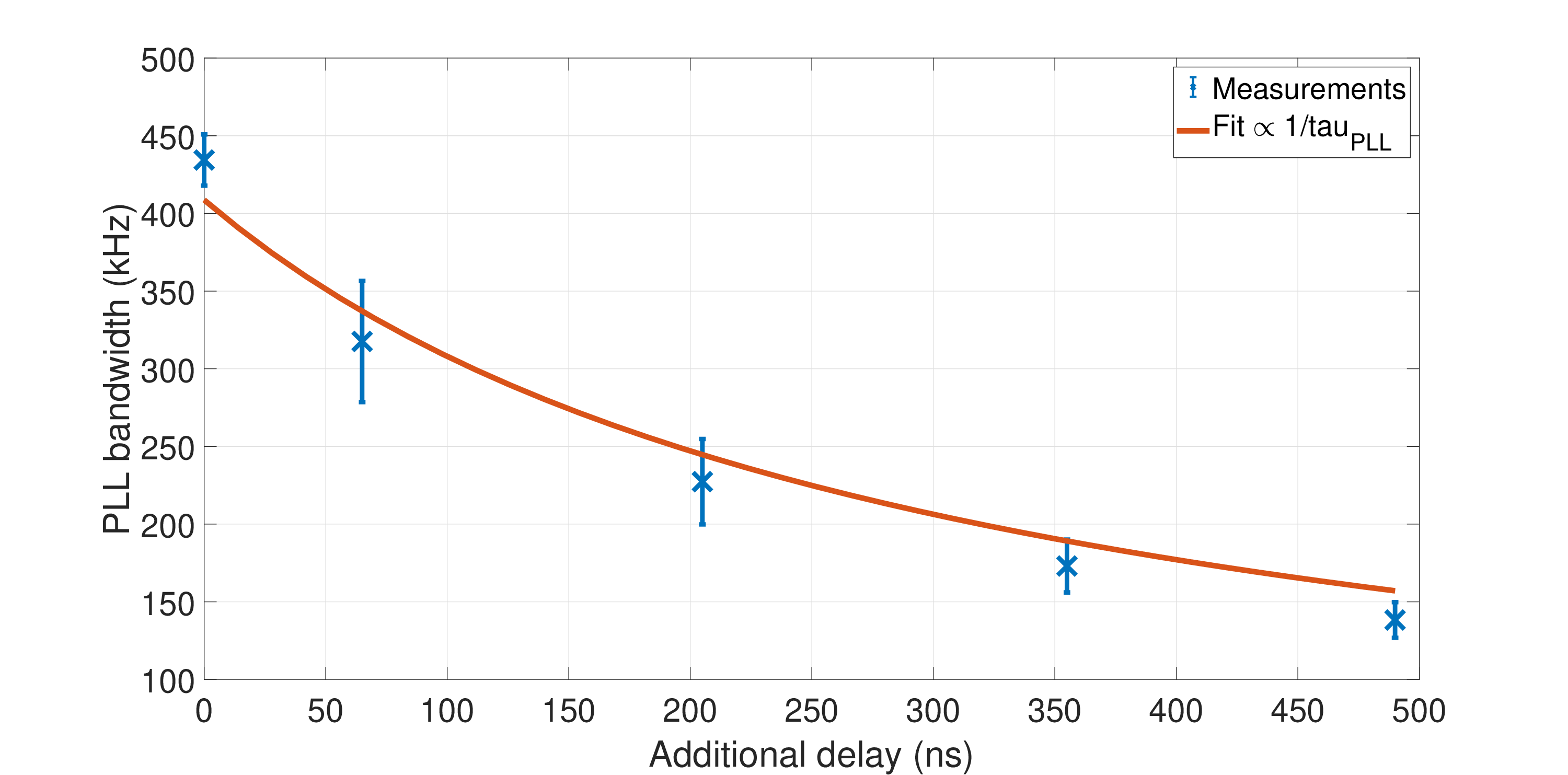}
\caption{Exploring the effect of adding coaxial delay lines to the feedback loop on the PLL bandwidth. The laser is operated at fixed frequency and the PLL bandwidth is retrieved from the beatnote spectra (analog to  Fig.~\ref{fig:psd_fb_ff}). The experimental points are fitted using Eq.~\ref{eq:pll_bw} and a value $\tPLL = 306 \pm 12$~ns is found.}
\label{fig:tauPLL}
\end{figure}

We experimentally examine the influence of the total loop delay on the feedback loop performance. The laser is operated at fixed frequency and the feedback bandwidth is retrieved from the beatnote spectra (analog to  Fig.~\ref{fig:psd_fb_ff}). Several values of the PLL delay are explored by adding coaxial delay lines to the existing setup, leading to a total PLL length $\tPLL+\tau_\mathrm{coax}$. The measured bandwidths are shown in Figure~\ref{fig:tauPLL} and show a $1/(\tPLL+\tau_\mathrm{coax})$ scaling law, in agreement with Eq.~\ref{eq:pll_bw}. By fitting the measurements with the model, we derive $\tPLL = 306 \pm 12$~ns.

\subsection{MZI delay influence on instantaneous frequency measurement}
Although the MZI delay only marginally affects the PLL transfer function and bandwidth, it actually influences the laser frequency measurement precision, because it is involved in the derivation of the error signal. In the following we show that the ideal MZI delay results from a compromise between precision and sensitivity to detection noise.

\paragraph{Instantaneous frequency accuracy}
We first analyze how the MZI delay comes into play in the laser instantaneous frequency derivation. This derivation starts with the analytical representation of the beatnote signal at the output of the MZI, whose complex phase is $2 \pi \fAOM t + \phi(t-\tMZI) - \phi(t)$. After removing the linear term due to the frequency shift $2 \pi \fAOM t$, one is left with the laser phase variation $\phi(t-\tMZI) - \phi(t)$, from which the laser instantaneous frequency is derived:
\begin{equation}
    f(t)=-\frac{1}{2\pi} \frac{d\Phi}{dt}\simeq-\frac{1}{2\pi}  \frac{\Phi(t)-\Phi(t-\tMZI)}{\tMZI}
    \label{eq:Taylor}
\end{equation}
This estimation of the instantaneous frequency is not perfectly accurate. For example, in the case of a linear frequency chirp $\phi(t) = \pi r t^2$, Eq.~\ref{eq:Taylor} leads to a $\frac 1 2 r\tMZI$ inaccuracy of the instantaneous frequency estimation. This inaccuracy increases with the MZI delay~\cite{ahn2007analysis}.

\paragraph{Sensitivity to detection noise}
On the other hand, shorter delays increase the sensitivity to detection noise. When the laser is perfectly stable with no frequency drift ($f(t)=0$), the beatnote signal captured by the photodiode is a sinusoid oscillating at the AOM frequency: $V(t) = V_{0} (1 + \sin(2 \pi \fAOM t)~)$. In the presence of a monochromatic detection noise at frequency $F$, the voltage becomes $V(t)=V_{0} \left(1 + \sin(2 \pi \fAOM t )\right) + A \sin(2 \pi Ft)$.
Once the constant term $V_{0}$ is removed, its instantaneous phase $\psi(t)$ is obtained as the complex argument of the analytic representation of $V(t)$~\cite{ahn2007analysis}.
Assuming $A \ll V_{0}$, one obtains:
\begin{equation}
    \psi(t) = 2\pi \fAOM t + \frac{A}{V_{0}} \sin \left( 2\pi (F-\fAOM)t \right)
\end{equation}
After removing the linear phase coming from the fixed frequency shift in the MZI ($2 \pi \fAOM t$), the remaining term is multiplied by $\frac{1}{2 \pi \tMZI}$ to derive the instantaneous frequency (according to Equation~\ref{eq:Taylor}):
\begin{equation}
    f(t) = \frac{\frac{A}{V_{0}}}{2\pi\tMZI } \sin \left(2\pi(F-\fAOM)t \right)
\end{equation}
Therefore the instantaneous frequency derivation converts the detection noise into instantaneous frequency measurement noise. 
The variance of such  frequency fluctuations is given by:
\begin{equation}
    \sigma_\mathrm{det}^2(F) = \frac{(\frac{A}{V_{0}})^2}{8 \pi^2 \tMZI^2}
\end{equation}
This frequency-independent, $1/\tMZI^2$ scaling law shows that a long MZI delay minimizes the influence of electronic noise.

\begin{figure}[t]
\centering \includegraphics[width=8.5cm]{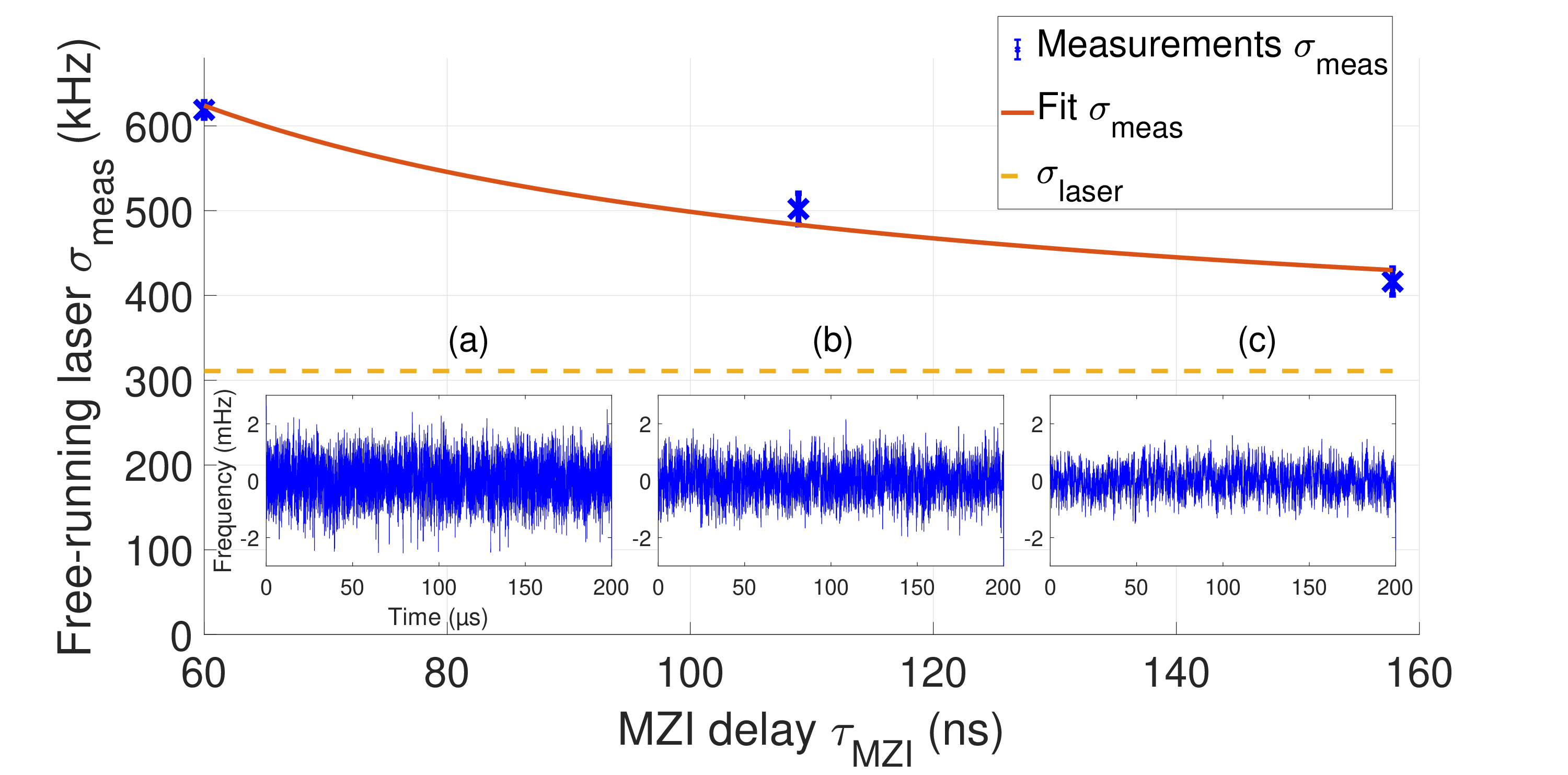}
\caption{Influence of the MZI delay on the occurence of detection noise. The RMS frequency error is measured over a $200~\mu$s interval for different MZI delays. The laser is operated in free-running mode (no feedback or feed-forward) at fixed frequency. The measurements are fitted using Eq.~\ref{eq:sommevariances} with $\sigma_\mathrm{laser} = 310$~kHz. Insets: corresponding instantaneous frequency measurements: (a) $\tMZI=60$ ns (b) $108.9$ ns and (c) $158$ ns.}
\label{fig:mesures_h0_f_tau_avec_DSP}
\end{figure}

Generalizing to a white detection noise, the power spectral density of the apparent laser frequency noise should also scale as $1/\tMZI^2$. We verify this by measuring the free-running laser frequency variance using different MZI delay lengths (see Fig.~\ref{fig:mesures_h0_f_tau_avec_DSP}). We observe that the frequency noise variance is compatible with the sum of two terms:
\begin{equation}
    \sigma^{2} = \sigma_\mathrm{laser}^{2} + \sigma_\mathrm{det}^{2}
    \label{eq:sommevariances}
\end{equation}
where $\sigma_\mathrm{laser}^2$ is the native laser frequency noise variance and $\sigma_\mathrm{det}^2\propto 1/\tMZI^2$ corresponds to the contribution of detection noise.

\paragraph{Conclusion}
As a result, although the MZI delay has little influence on the PLL bandwidth, it has a strong impact on the accuracy of the instantaneous derivation on the one hand, and on the transduction of electronic noise onto the instantaneous frequency measurement.
We finally choose a delay of $60$~ns, as a compromise between precision and sensitivity to detection noise.

\section{Appendix: Feed-forward}
\label{app:feedforward}

In this section we discuss the feed-forward theoretical transfer function and the influence of the EOM delay. We show that the optimal delay can be determined by minimizing the laser linewidth.

\subsection{Feed-forward delay}

\begin{figure}[t]
\centering \includegraphics[width=8.5cm]{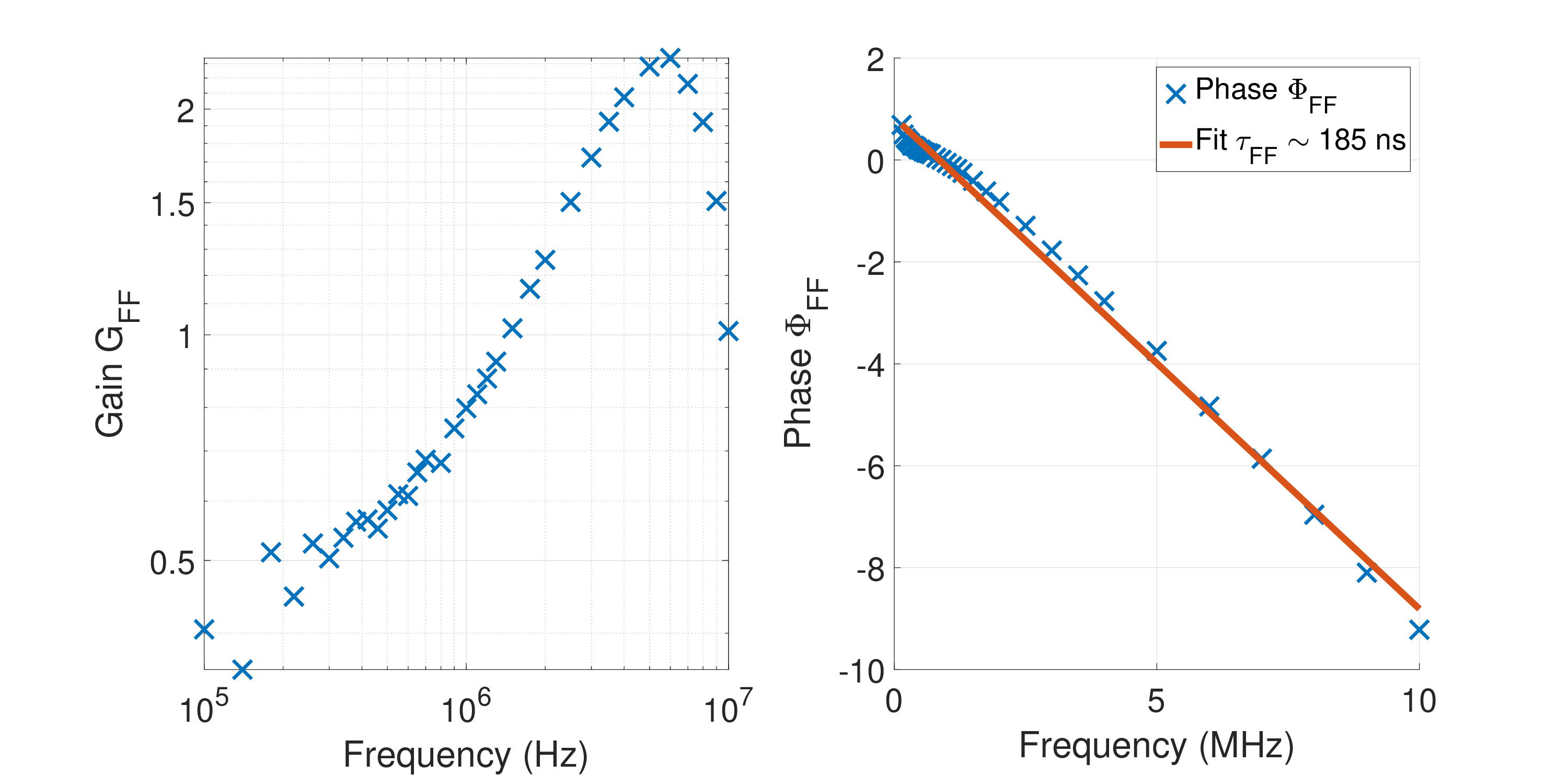}
\caption{Feed-forward transfer function. The feed-forward exhibits a high-pass behaviour due to the derivator behavior of the EOM. Nevertheless, a cutoff is observed above $5$~MHz due to the MZI response, determining the FF bandwidth. The ideal length of the fibered delay line $\tEOM$ is derived from the slope of the transfer function phase (see Eq.~\ref{eq:PhiFF}).}
\label{fig:fit_tau_EOM_optimise}
\end{figure}

The buildup of the feed-forward correction signal is subject to a time delay $\tFF$ owing to the various optical and electronic components of the feed-forward setup:
\begin{equation}
    \tFF = \frac{-1}{2\pi} \frac{\partial \Phi_\mathrm{FF}}{\partial f}
    \label{eq:tauFF}
    \end{equation}
where
\begin{equation}
\Phi_\mathrm{FF}=\arg\left[ \FMZI(f) F_{\Phi}(f) \FFF (f) F_\mathrm{lp}(f) \FEOM (f) \right]
    \label{eq:PhiFF}
\end{equation}
$\FMZI$, $F_\Phi$, $\FFF$, $F_\mathrm{lp}$, $\FEOM$ are the transfer functions of the MZI, the phase comparator, the feed-forward PI filter, the low-pass filter, and the electro-optic modulator, respectively. Some of these transfer functions can be measured independently, namely $F_\Phi$, $\FFF$, and $F_{lp}$. Measuring the transfer function of an EOM requires a means to measure an instantaneous frequency in the optical domain. This is why it is most convenient to measure the combined transfer functions of the EOM and the MZI: $\FMZI(f)\FEOM(f)$. We are finally able to compute the product of all these transfer functions, and we plot them in Fig.~\ref{fig:fit_tau_EOM_optimise}. We find that the feed-forward signal is high-pass filtered due to the derivator behaviour of the EOM. It exhibits a linear phase $\Phi_\mathrm{FF}(f)$, indicating an effective group delay $\tFF=183$~ns. This delay corresponds to the sum of all individual group delays of each element of the feed-forward correction, which are specified in Table~\ref{tab:ff_group_delays}.

\begin{table}[t]
\centering
\begin{tabular}{ccc}
\hline \hline
Component & Transfer function & Group delay\\
\hline
EOM & $\FMZI(f)\FEOM(f)$ & $90$~ns \\
Mixer & $F_\Phi(f)$ & $50$~ns \\
LB1005 & $F_\mathrm{PI}(f)$ & $40$~ns \\
Low-pass filter & $F_\mathrm{lp}(f)$ & $15$~ns \\
\hline \hline
 \end{tabular}
\caption{Effective group delays induced by each element of the feed-forward correction.}
\label{tab:ff_group_delays}
\end{table}

\subsection{Influence of EOM delay mismatch}
The feed-forward optimally corrects the laser frequency deviations when the delay $\tFF$ experienced by the correction signal to reach the EOM matches the physical delay $\tEOM$ experienced by the laser beam to reach that same device. In this section we study the influence of a feed-forward delay mismatch $\dtEOM$ on the laser linewidth. To that end we unfold a simple theoretical approach.

Let us first assume that the laser source exhibits a sinusoidal frequency fluctuation $\epsilon_0 (t) = \Delta F \sin(2 \pi F t)$. We consider that the feed-forward correction signal $G \epsilon_0(t-\dtEOM)$ is applied to the EOM with a temporal mismatch $\dtEOM$, leading to a corrected frequency fluctuation:
\begin{equation}
    \epsilon(t) = \Delta F \left( \sin \left[2 \pi F t \right] - G \sin \left[2 \pi F (t-\dtEOM) \right] \right)
\end{equation}
The frequency fluctuation variance $\sigma^{2} (F) $, defined as: $\sigma^{2} =  \langle \epsilon(t)^2 \rangle$ is given by:
\begin{equation}
    \sigma^{2}(F) = \Delta F^2 \left( \frac{G^2+1}{2} - G \cos (2\pi F\dtEOM) \right)
\end{equation}

\begin{figure}[t]
\centering \includegraphics[width=8.5cm]{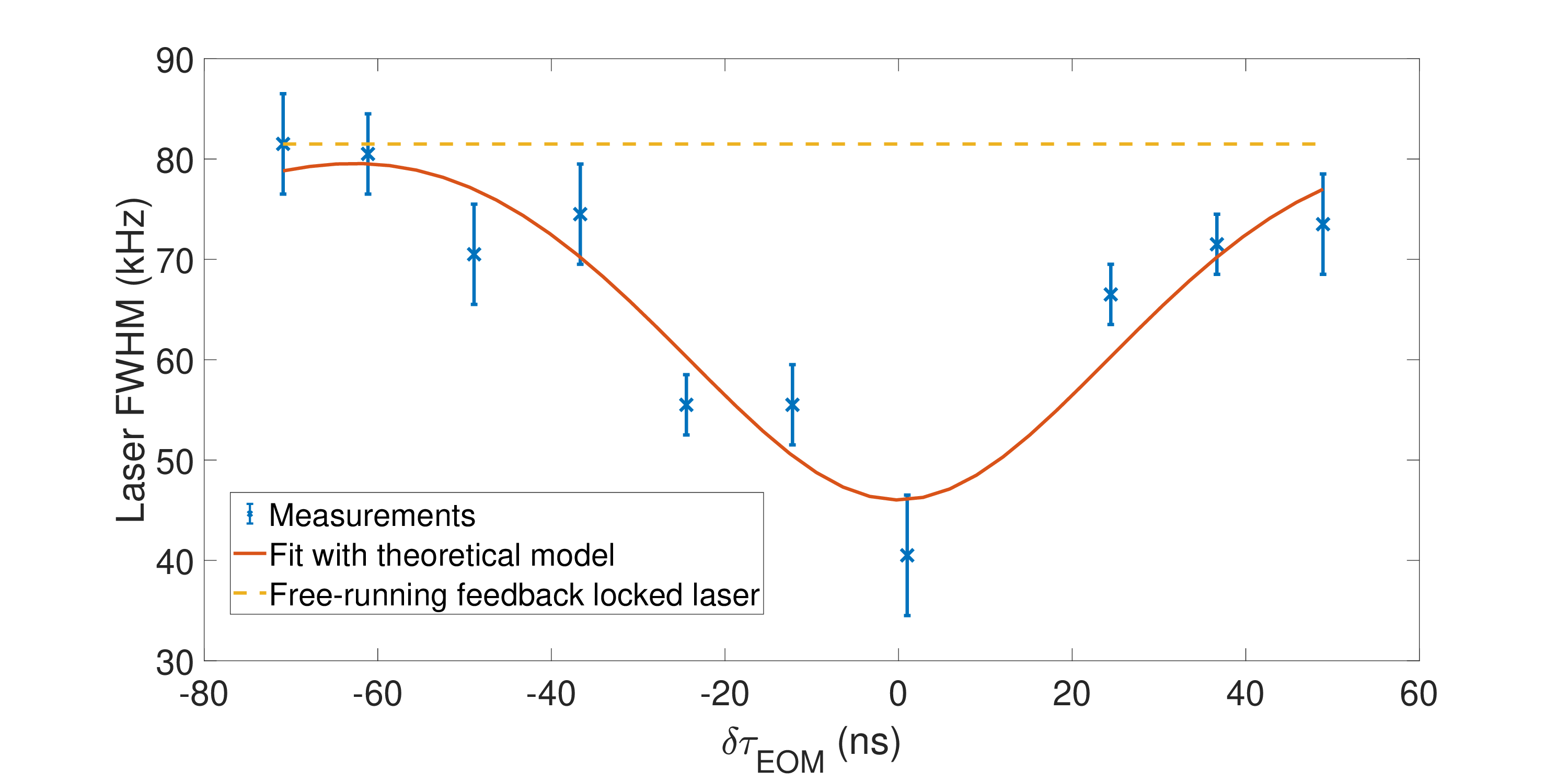}
\caption{Influence of a feed-forward delay mismatch $\dtEOM$ on the laser linewidth ($10$~ms integration time). The latter is measured with a stabilized optical frequency discriminator (SilentSys). Zero $\dtEOM$ corresponds to the optimised FF delay.}
\label{fig:laser_linewidth_EOMdelay}
\end{figure}

We now consider that the laser is affected by a white frequency noise, \emph{ie} the perturbation is uniform for all frequencies. The laser linewidth is proportional to the total variance of the laser frequency fluctuations~\cite{elliott1982extracavity}:
\begin{equation}
    \sigma^{2}_{tot}  = \frac{1}{F_{max} - F_{min}} \int_{F_{min}}^{F_{max}} \sigma^{2} (F) dF
\end{equation}
where $F_{max}$ and $F_{min}$ are the maximum and minimum frequency addressed by the feed-forward correction. Assuming $F_{min} \ll F_{max}$, one obtains:
\begin{equation}
\sigma^{2}_{tot} =\Delta F ^{2} \left( \frac{G^2+1}{2} - G \frac{\sin(2 \pi \dtEOM F_{max})}{2 \pi \dtEOM F_{max}}  \right)
  \label{eq:sigma_tot}
\end{equation}
This expression shows that the laser linewidth is minimal when $\dtEOM=0$, in agreement with empirical measurements reported in~\cite{lintz2017note}. It also highlights that finding such a minimum is easier when the feed-forward bandwidth is reduced. We verify this behaviour by measuring the DBR laser linewidth with a range of values for the EOM delay around the optimal $\tEOM$, using optical fibers with different lengths (see Figure~\ref{fig:laser_linewidth_EOMdelay}). Fitting the experimental data with expression~\ref{eq:sigma_tot}, we find the actual feed-forward correction bandwidth $F_{max} = 5.7 \pm 0.6$~MHz.

\begin{backmatter}
\bmsection{Funding}
The authors acknowledge support from the French National Research Agency (ANR)
through the ATRAP project (ANR-19-CE24-0008). This work has received support under the program “Investissements d’Avenir” launched by the French Government.

\bmsection{Acknowledgments}
The authors are grateful to Michel Lintz and Vincent Crozatier for helpful discussions.

\bmsection{Disclosures}The authors declare no conflicts of interest.

\bmsection{Data availability} Data underlying the results presented in this paper are not publicly available at this time but may be obtained from the authors upon reasonable request.

\end{backmatter}


\begin{thebibliography}{10}
\newcommand{\enquote}[1]{``#1''}

\bibitem{Gazizov2022lowpixel}
I.~Gazizov, S.~Zenevich, and A.~Rodin, \enquote{Low-pixel-count imaging fmcw
  lidar,} {\protect\JournalTitle{Appl. Opt.}} \textbf{61}, 9241--9246 (2022).

\bibitem{zhang2019laser}
X.~Zhang, J.~Pouls, and M.~C. Wu, \enquote{Laser frequency sweep linearization
  by iterative learning pre-distortion for fmcw lidar,}
  {\protect\JournalTitle{Opt. Express}} \textbf{27}, 9965--9974 (2019).

\bibitem{kranendonk2005modeless}
L.~A. Kranendonk, R.~J. Bartula, and S.~T. Sanders, \enquote{Modeless operation
  of a wavelength-agile laser by high-speed cavity length changes,}
  {\protect\JournalTitle{Optics Express}} \textbf{13}, 1498--1507 (2005).

\bibitem{babbitt2014spectral}
W.~R. Babbitt, Z.~W. Barber, S.~H. Bekker, \emph{et~al.}, \enquote{From
  spectral holeburning memory to spatial-spectral microwave signal processing,}
  {\protect\JournalTitle{Laser Physics}} \textbf{24}, 094002 (2014).

\bibitem{kinos2021roadmap}
A.~Kinos, D.~Hunger, R.~Kolesov, \emph{et~al.}, \enquote{Roadmap for rare-earth
  quantum computing,} {\protect\JournalTitle{arXiv preprint arXiv:2103.15743}}
  (2021).

\bibitem{levin2002mode}
L.~Levin, \enquote{Mode-hop-free electro-optically tuned diode laser,}
  {\protect\JournalTitle{Optics letters}} \textbf{27}, 237--239 (2002).

\bibitem{crozatier2006phase}
V.~Crozatier, G.~Gorju, F.~Bretenaker, \emph{et~al.}, \enquote{Phase locking of
  a frequency agile laser,} {\protect\JournalTitle{Applied Physics Letters}}
  \textbf{89} (2006).

\bibitem{li2022integrated}
M.~Li, L.~Chang, L.~Wu, \emph{et~al.}, \enquote{Integrated {Pockels} laser,}
  {\protect\JournalTitle{Nature communications}} \textbf{13}, 5344 (2022).

\bibitem{nagarajan1999high}
R.~Nagarajan and J.~E. Bowers, \enquote{High-speed lasers,}
  {\protect\JournalTitle{Semiconductor Lasers I}} pp. 177--290 (1999).

\bibitem{westbrook1984monolithic}
L.~Westbrook, A.~Nelson, P.~Fiddyment, and J.~Collins, \enquote{Monolithic 1.5
  $\mu$m hybrid {DFB/DBR} lasers with 5 nm tuning range,}
  {\protect\JournalTitle{Electronics letters}} \textbf{23}, 957--959 (1984).

\bibitem{pan1989modulation}
X.~Pan, H.~Olesen, and B.~Tromborg, \enquote{Modulation characteristics of
  tunable {DFB/DBR} lasers with one or two passive tuning sections,}
  {\protect\JournalTitle{IEEE Journal of Quantum Electronics}} \textbf{25},
  1254--1260 (1989).

\bibitem{zheng2007determination}
W.~Zheng and G.~Taylor, \enquote{Determination of the photon lifetime for dfb
  lasers,} {\protect\JournalTitle{IEEE Journal of Quantum Electronics}}
  \textbf{43}, 295--302 (2007).

\bibitem{yi2021frequency}
W.~Yi, Z.~Li, Z.~Zhou, \emph{et~al.}, \enquote{Frequency-modulated chirp
  signals for single-photodiode based coherent lidar system,}
  {\protect\JournalTitle{Journal of Lightwave Technology}} \textbf{39},
  4661--4670 (2021).

\bibitem{laroche2008serrodyne}
M.~Laroche, C.~Bartolacci, G.~Lesueur, \emph{et~al.}, \enquote{Serrodyne
  optical frequency shifting for heterodyne self-mixing in a
  distributed-feedback fiber laser,} {\protect\JournalTitle{Optics letters}}
  \textbf{33}, 2746--2748 (2008).

\bibitem{poberezhskiy2005serrodyne}
I.~Y. Poberezhskiy, B.~Bortnik, J.~Chou, \emph{et~al.}, \enquote{Serrodyne
  frequency translation of continuous optical signals using ultrawide-band
  electrical sawtooth waveforms,} {\protect\JournalTitle{IEEE journal of
  quantum electronics}} \textbf{41}, 1533--1539 (2005).

\bibitem{satyan2009precise}
N.~Satyan, A.~Vasilyev, G.~Rakuljic, \emph{et~al.}, \enquote{Precise control of
  broadband frequency chirps using optoelectronic feedback,}
  {\protect\JournalTitle{Optics express}} \textbf{17}, 15991--15999 (2009).

\bibitem{kervella2014laser}
G.~Kervella, J.~Maxin, M.~Faugeron, \emph{et~al.}, \enquote{Laser sources for
  microwave to millimeter-wave applications,} {\protect\JournalTitle{Photonics
  Research}} \textbf{2}, B70--B79 (2014).

\bibitem{lihachev2023frequency}
G.~Lihachev, A.~Bancora, V.~Snigirev, \emph{et~al.}, \enquote{Frequency agile
  photonic integrated external cavity laser,} {\protect\JournalTitle{arXiv
  preprint arXiv:2303.00425}}  (2023).

\bibitem{li2020nonlinear}
P.~Li, T.~Yang, J.~Yang, \emph{et~al.}, \enquote{The nonlinear tuning technique
  of a {DFB} laser using frequency predistortion procedures,} in
  \emph{Terahertz, RF, Millimeter, and Submillimeter-Wave Technology and
  Applications XIII,}  vol. 11279 (SPIE, 2020), pp. 250--255.

\bibitem{cao2021highly}
X.~Cao, K.~Wu, C.~Li, \emph{et~al.}, \enquote{Highly efficient iteration
  algorithm for a linear frequency-sweep distributed feedback laser in
  frequency-modulated continuous wave lidar applications,}
  {\protect\JournalTitle{JOSA B}} \textbf{38}, D8--D14 (2021).

\bibitem{qin2015coherence}
J.~Qin, Q.~Zhou, W.~Xie, \emph{et~al.}, \enquote{Coherence enhancement of a
  chirped dfb laser for frequency-modulated continuous-wave reflectometry using
  a composite feedback loop,} {\protect\JournalTitle{Optics letters}}
  \textbf{40}, 4500--4503 (2015).

\bibitem{roos2009ultrabroadband}
P.~A. Roos, R.~R. Reibel, T.~Berg, \emph{et~al.}, \enquote{Ultrabroadband
  optical chirp linearization for precision metrology applications,}
  {\protect\JournalTitle{Optics letters}} \textbf{34}, 3692--3694 (2009).

\bibitem{li2022linear}
P.~Li, Y.~Zhang, and J.~Yao, \enquote{Linear frequency swept laser source with
  high swept slope based on digital optical phase-locked loop,}
  {\protect\JournalTitle{Optics Communications}} \textbf{525}, 128860 (2022).

\bibitem{greiner1998laser}
C.~Greiner, B.~Boggs, T.~Wang, and T.~Mossberg, \enquote{Laser frequency
  stabilization by means of optical self-heterodyne beat-frequency control,}
  {\protect\JournalTitle{Optics letters}} \textbf{23}, 1280--1282 (1998).

\bibitem{lintz2017note}
M.~Lintz, D.-H. Phung, J.-P. Coulon, \emph{et~al.}, \enquote{Note: Efficient
  diode laser line narrowing using dual, feed-forward+ feed-back laser
  frequency control,} {\protect\JournalTitle{Review of Scientific Instruments}}
  \textbf{88} (2017).

\bibitem{cheng2022feed}
Y.~S. Cheng, B.~Szutor, and D.~T. Reid, \enquote{Feed-forward stabilization of
  a single-frequency, diode-pumped {Pr:YLF-Cr:LiCAF} laser operating at 813.42
  nm,} {\protect\JournalTitle{Optics Express}} \textbf{30}, 42902--42911
  (2022).

\bibitem{zhang2019research}
J.~Zhang, C.~Gao, M.~Xue, and R.~Liu, \enquote{Research on frequency modulation
  character of the current driven dfb semiconductor laser,}
  {\protect\JournalTitle{Modern Physics Letters B}} \textbf{33}, 1850422
  (2019).

\bibitem{ahn2007analysis}
T.-J. Ahn and D.~Y. Kim, \enquote{Analysis of nonlinear frequency sweep in
  high-speed tunable laser sources using a self-homodyne measurement and
  hilbert transformation,} {\protect\JournalTitle{Applied optics}} \textbf{46},
  2394--2400 (2007).

\bibitem{funabashi2004recent}
M.~Funabashi, H.~Nasu, T.~Mukaihara, \emph{et~al.}, \enquote{Recent advances in
  {DFB} lasers for ultradense {WDM} applications,} {\protect\JournalTitle{IEEE
  Journal of selected topics in quantum electronics}} \textbf{10}, 312--320
  (2004).

\bibitem{aflatouni2012wideband}
F.~Aflatouni and H.~Hashemi, \enquote{Wideband tunable laser phase noise
  reduction using single sideband modulation in an electro-optical feed-forward
  scheme,} {\protect\JournalTitle{Optics Letters}} \textbf{37}, 196--198
  (2012).

\bibitem{didomenico2010simple}
G.~Di~Domenico, S.~Schilt, and P.~Thomann, \enquote{Simple approach to the
  relation between laser frequency noise and laser line shape,}
  {\protect\JournalTitle{Applied optics}} \textbf{49}, 4801--4807 (2010).

\bibitem{vonbandel2016time}
N.~Von~Bandel, M.~Myara, M.~Sellahi, \emph{et~al.}, \enquote{Time-dependent
  laser linewidth: beat-note digital acquisition and numerical analysis,}
  {\protect\JournalTitle{Optics Express}} \textbf{24}, 27961--27978 (2016).

\bibitem{stephan2005laser}
G.~M. St{\'e}phan, T.~Tam, S.~Blin, \emph{et~al.}, \enquote{Laser line shape
  and spectral density of frequency noise,} {\protect\JournalTitle{Physical
  Review A}} \textbf{71}, 043809 (2005).

\bibitem{blanchard1976pll}
A.~{Blanchard}, \emph{{Phase-locked loops: Application to coherent receiver
  design}} (1976).

\bibitem{elliott1982extracavity}
D.~S. Elliott, R.~Roy, and S.~J. Smith, \enquote{Extracavity laser band-shape
  and bandwidth modification,} {\protect\JournalTitle{Phys. Rev. A}}
  \textbf{26}, 12--18 (1982).

\end{thebibliography}

\end{document}